# A Matter of Principle: The Principles of Quantum Theory, Dirac's Equation, and Quantum Information

## Arkady Plotnitsky[a]

[a] *Purdue University, West Lafayette, IN 47907 e-mail: plotnits@.purdue.edu*

**Abstract.** This article is concerned with the role of fundamental principles in theoretical physics, especially quantum theory. The fundamental principles of relativity will be addressed as well, in view of their role in quantum electrodynamics and quantum field theory, specifically Dirac's work, which, in particular Dirac's derivation of his relativistic equation of the electron from the principles of relativity and quantum theory, is the main focus of this article. I shall also consider Heisenberg's earlier work leading him to the discovery of quantum mechanics, which inspired Dirac's work. I argue that Heisenberg's and Dirac's work was guided by their adherence to and their confidence in the fundamental principles of quantum theory. The final section of the article discusses the recent work by G. M. D'Ariano and coworkers on the principles of quantum information theory, which extend quantum theory and its principles in a new direction. This extension enabled them to offer a new derivation of Dirac's equations from these principles alone, without using the principles of relativity.

> I am working out a quantum theory about it for it is really most tantalizing state of affairs.
>
> —James Joyce, *Finnegans Wake*

## 1. Introduction

In his paper introducing his famous relativistic equation for the electron, Dirac commented on a major difficulty that his equation inherited from the Klein-Gordon equation, even though he saw his equation as a major advance. In particular, unlike the Klein-Gordon equation, Dirac's equation enabled him to answer, for the relativistic electron, the following question, arguably the main question of quantum theory: "What is the probability of *any dynamical variable* at any specified time having a value laying between any specified limits, when the system is represented by a given wave function $\psi_n$?" Dirac clarified that the Klein-Gordon theory "can answer such questions if they refer to the position of the electron … but not if they refer to its momentum, or angular momentum, or any other dynamic variable" [1, pp. 611-612; emphasis added]. However, as Dirac explains, there was a major problem equally affecting both theories:

[Either equation] refers equally well to an electron with charge e as to one with charge – e. If one considers for definitiveness the limiting case of large quantum numbers one would find that some of the solutions of the wave equation are wave packets moving in the way a particle of – e would move on the classical theory, while others are wave packets moving in the way a particle with charge e would move classically. For this second class of solutions W has a negative value. One gets over the difficulty on the classical theory by arbitrarily excluding those solutions that have a negative W. One cannot do this on the quantum theory, since in general a perturbation will cause transitions from state with W positive to states with W negative. Such a transition would appear experimentally as the electron suddenly changes its charge from – e to e, a phenomenon which has not been observed. The true relativi[stic] wave equation should thus be such that its solutions split up into two non-combining sets, referring respectively to the charge – e and the charge e.

In the present paper we shall only be concerned with the removal of [the Klein-Gordon equation's inability to predict the probability of all dynamical variables of the system  represented by a given wave function]. The resulting theory [embodied in Dirac's equation] is therefore still only an approximation, but it appears to be good enough to account for all the duplexity phenomena without arbitrary assumptions. [1, p. 612]

The difficulty does not appear in low-energy regimes or rather it disappears at the low energy limit, because Dirac's equation converts into Schrödinger's equation, where this problem does not arise. In commenting on the problem in his 1929 Chicago lectures, *The Physical Principles of the Quantum Theory*, published in 1930 [2], Heisenberg reprised Dirac's assessment and even amplified it by saying that Dirac's "theory is certainly wrong." Heisenberg added an intriguing twist. He said: "The classical theory could eliminate this difficulty by arbitrarily excluding the one sign, but this is not possible to do according to *the principles of quantum theory*. Here spontaneous transitions may occur to the states of negative value of energy $E$; as these have never been observed, *the theory is certainly wrong*. Under these conditions it is very remarkable that the positive energy-levels (at least in the case of one electron) coincide with those actually observed" [2, p. 102; emphasis added]. As became apparent a few years later, Dirac's theory proved to be better than it appeared, even to its creator, at the time of its introduction. Indeed, it has proven to be correct. It was our understanding of nature in high-energy quantum regimes that was deficient. That a perturbation may cause transitions from a state with positive $E$ to a state with negative $E$ is a common occurrence that appears experimentally as an electron spontaneously changing its negative charge into a positive one and becoming a positron. Antimatter was staring into Dirac's eyes. It took, however, a few years to realize that it was antimatter and that this type of transition (eventually understood in terms of the creation and annihilation of particles and virtual particle formation) defines high-energy quantum regimes. This is one of the great stories of quantum theory.

The main concern of this paper, however, is another facet of Heisenberg's thinking transpiring here, to which I referred above as "an intriguing twist" of his assessment. This facet was equally found in Dirac's thinking. While Heisenberg and Dirac had their doubts concerning Dirac's theory itself, neither appears to have doubted "the *principles* of quantum theory," which, while physical in nature, were also given a proper mathematical expression in the formalism of the theory. This confidence and their emphasis on the role of these principles are reflected in the titles of Heisenberg's book, just cited, and Dirac's *The Principles of Quantum Mechanics,* published around the same time, in 1930 [3]. The term "principle" requires a proper explanation to be offered in Section 2. Suffice it to say for the moment that, as it was understood or (they do not really define it) at least practiced by Heisenberg and Dirac, and as it will be understood here, a *principle* is more than only a postulate expressing a fundamental assumption concerning how one should understand nature, although a principle may involve such a postulate. I shall explain the term postulate in Section 2. A principle also serves as guidance for a chain of reasoning in building a physical theory, making it what A. Einstein called "a principle theory," a concept also discussed in Section 2 [4]. It might have been a matter of replacing Dirac's theory with a different theory, based on these principles, but not replacing these principles themselves (because they had demonstrated their validity and efficacy in quantum mechanics), but only refining them and introducing new ones in the same spirit. As it happened, Dirac's theory has proven to be correct, although far from the end of the story of quantum electrodynamics and quantum field theory, which continue to conform to the principles in question. Heisenberg and Dirac were led to their momentous discoveries of, respectively, quantum mechanics and quantum electrodynamics by their confidence in and adherence to the fundamental principles of quantum theory, and by the discoveries of some of these principles, in the first place.

This article will consider the roles of fundamental principles and principle thinking in quantum theory, including thinking that leads to the invention of new such principles, which is, I would argue, one of the ultimate achievements of theoretical thinking in any field.[1] This subject has been uncommon in recent discussions of quantum foundations, dominated by more formal (mathematical or logical) aspects of quantum theory.[2] Dirac's work, most especially his 1928 discovery of his relativistic equation for the electron is my main example of principle thinking in quantum theory, although I shall also consider Heisenberg's principle thinking, which led him to his discovery of quantum mechanics. This thinking inspired much of Dirac's work, including that leading to his discovery of his equation. I shall offer a few brief reflections on the role of fundamental principles of quantum theory in subsequent developments of quantum field theory, leading to the Standard Model of elementary particle physics. Particular attention, however, will be given to the recent work by G. M. D'Ariano and coworkers, grounded in the principles of quantum theory as quantum information theory [9, 10, 11]. This rethinking allows them to derive Dirac's equation from the principles of quantum information alone, rather than, as Dirac did, from combining the principles of quantum theory and (special) relativity [11].

As just stated, rethinking fundamental principles is crucial to theoretical physics. W. Pauli stressed this point in his assessment of Einstein's work on quantum theory:

> If new features of the phenomena of nature are discovered that are incompatible with the system of theories assumed at that time, the question arises, which of the known principles used in the description of nature are general enough and which have to be modified or abandoned. The attitude of different physicists to problems of this kind, which make strong demand on the intuition and tact of a scientist, depends to a large extent on the personal temperament of the investigator. In the case of *Planck*'s discovery of 1900 of the quantum of action [*h*] during the course of his famous investigations of the law of the black-body radiation, it was clear that the law of conservation of energy and momentum and *Boltzmann*'s principle connecting entropy and probability were two pillars sufficiently strong to stand unshaken by the development resulting from the new discovery. It was indeed *the faithfulness to these principles* which enabled *Planck* to introduce the new constant *h*, the quantum of action, into his statistical theory of the thermodynamic equilibrium of radiation.
>
> The original investigation of *Planck*, however, had treated with a certain discretion the question whether the new "quantum-hypothesis" implies the necessity of changing the laws of microscopic phenomena themselves independent of statistical applications, or whether one had to use only an improvement of the statistical methods to enumerate equally probable states. In any case, the tendency towards a compromise between the older ideas of physics, now called the "classical" ones, and the quantum theory was always favored by *Planck*, both in his earlier and later work on the subject, although affirmation of such a possibility was to diminish considerably the significance of his own discovery.
>
> Such considerations formed the background on *Einstein*'s first paper on quantum theory … , which was

---

[1] Throughout this article, unless specified otherwise, by quantum *theory* I refer to the standard versions of quantum mechanics, quantum electrodynamics, and quantum field theory, especially their mathematical structures, rather than alternative theories of quantum phenomena, such as Bohmian theories, for example. By quantum phenomena, I refer to those observed physical phenomena in considering which Planck's constant, *h*, must be taken into account; and by quantum objects, I refer to those entities in nature that, through their interactions with measuring instruments, are responsible for the appearance of quantum phenomena.

[2] Among exceptions are A. Zeilinger's article [5], J. Bub's article on quantum mechanics as a principle theory on Einstein's definition [6], an earlier approach to Heisenberg's discovery of quantum mechanics by the present author [7, pp. 9-16), and most recently, R. M. Ungar and L. Smolin's book, which builds on Smolin's earlier work [6]. The principles grounding Smolin's argument are, however, not the principles of quantum theory considered in this article. Indeed, most of his key principles, beginning with Leibniz's principle of sufficient reason, which grounds Smolin's argument, are in conflict with the key principles of quantum theory advocated here. (Smolin's argumentation is in conflict with some of the principles of relativity, both special and general, as well.) There is some overlap. The gauge-invariance principle, extensively used by Smolin, is consistent with the principles advocated here and is especially important in quantum field theory. Also, Smolin's view of mathematics and its role in physics is in accord with the present argument and the overall (non-Platonist) philosophical stance adopted in this article.

preced by his papers on the fundamental of statistical mechanics and accompanies, in the same year 1905, by his fundamental papers on the theory of the Brownian motion and the theory of relativity. [12, p. 86; emphasis added]

Later on Einstein also reflected, in explaining the (principle) nature of relativity theory, on the concept of (fundamental) principles and on "principle [physical] theories," as he called them, in juxtaposition to "constructive theories" [4]. Pauli does not explain what he means by principles, nor, as I said, do Heisenberg and Dirac in their books, their titles notwithstanding. Pauli, however, is right to argue that "it was indeed *the faithfulness to these principles* which enabled Planck to introduce the new constant *h*, the quantum of action, into his statistical theory of the thermodynamic equilibrium of radiation" (emphasis added). Heisenberg's and Dirac's great discoveries were equally enabled by their faithfulness to the principles of quantum theory and by their introduction of new such principles.

## 2. Principle Thinking in Theoretical Physics

This section explains the concept of principle, as it is to be understood in this article. While it has not always been expressly defined by the founding figures mentioned above, how they use the concept of principle could be given a cohesive sense, with which the present understanding of principle is consistent and from which it in part derives. Ultimately, these are this use and, in the first place, the invention of new principles that are most crucial, especially when it comes to the discovery of new theories in physics or elsewhere, a point not lost on Einstein, whose distinction between "constructive" and "principle" theories is my point of departure here [4].

First, I would like to define "axioms" and "postulates," again, as they will be used here, although these terms become germane to my argument only in Section 5. These terms are often used, in physics (mathematicians tend to be more careful), somewhat indiscriminately and interchangeably with each other or both with "principle," as a result obscuring substantive points at stake. Admittedly, it is difficult and perhaps impossible to entirely avoid overlapping between the concepts designated by these terms, but it is possible to sufficiently analytically separate these concepts, enabling a better understanding of their functioning and roles, and I shall attempt to do so here. Euclid, and, it appears, the ancient Greeks in general, distinguished between the "axioms" and the "postulates." Axioms were thought to be something manifestly self-evident, such as the first axiom of Euclid ("things equal to the same thing are also equal to each other"). A postulate, by contrast, is *postulated*, in the sense of "let us assume that … ," thus indicating more that one aims to proceed under this assumption and see what follows from it according to established logical rules (this is the same as proceeding from axioms), rather than claiming the postulate to be a self-evident truth. Euclid's postulates may be thought of as those assumptions that were necessary and sufficient to derive the truths of geometry, of some of which we might already be intuitively persuaded (e.g., "the first postulate: to draw a straight line from any point to any point"). The famous Fifth Postulate is a case in point. It defines Euclidean (flat) geometry alone, which also explains millennia of attempts to derive it as a theorem. In Kant's understanding of geometry, inspired by Euclid, axioms are analytic and postulates synthetic propositions. Keeping in mind further complexities potentially involved in geometry and beyond, I shall adopt this understanding of axioms and postulates. Given that my subject is physics rather than mathematics, I shall refer primarily to postulates, assumed on the basis of experimental evidence (as it stands now and hence potentially refutable) and often, but not always, grounding principles. In fact, it is not easy to speak of axioms in the sense just defined in fundamental physics, either relativistic or quantum. Next to nothing has the self-evidence of axioms, and most of the uses of the term "axiom" are in effect closer to that of "postulate" as just defined.

Einstein's distinction between "constructive" and "principle" theories represents two contrasting, although in practice often intermixed, ways of thinking in fundamental physics. Before I explain them, I would like to give an example, one of the earliest examples of the use of a principle in modern physics, Pierre de Fermat's "principle of least time," eventually developed (a number of figures were involved) into the principle of least action. Both principles involve postulates. Fermat used the principle to explain the so-called Snell law describing the refraction of light passing through a slab of glass. Fermat's principle was not necessary or even especially helpful at the time, because one could more easily use the Snell law in doing calculations in specific cases, while one would ultimately need the calculus of variations to do so from Fermat's principle and to give it a proper mathematical expression. However, Fermat's principle defined *how nature works*, and as such, it was profound and far-reaching, especially once it was developed into the principle of least action. The subsequent history of physics has demonstrated the profundity and power of this principle on many occasions, including in relativity and quantum mechanics. It played key roles in D. Hilbert's derivation of Einstein's equations of general relativity and E. Noether's proof of her celebrated theorems relating symmetry and conservation laws.

According to Einstein, "constructive theories" aim "to build up a picture of the more complex phenomena out of the materials of a relatively simple formal scheme from which they start out," which customarily implies that this simpler formal scheme describes, at least ideally and in principle, the ultimate underlying reality responsible for this phenomena [4, p. 228]. Einstein's example of a constructive theory in classical physics is the kinetic theory of gases, which "seeks to reduce mechanical, thermal, and diffusional processes to movements of molecules—i.e., to build them up out of the hypothesis of molecular motion," described by the laws of classical mechanics [4, p. 228]. This assumption was *in effect* abandoned by Planck in building his black body radiation theory, which led to the rise of quantum physics, although it was Einstein and not Planck (hence "in effect") who was the first to realize this incompatibility between Planck's quantum hypothesis and this underlying hypothesis of classical statistical physics [13]. One could, however, build quantum theory independently of this assumption or indeed independently of *constructing* "the materials of a relatively simple [underlying] formal scheme," but instead as a principle theory, as Heisenberg did in the case of matrix mechanics.

It contrast to constructive theories, principle theories, according to Einstein, "employ the analytic, not the synthetic, method. The elements which form their basis and starting point are not hypothetically constructed but empirically discovered ones, general characteristics of natural processes, principles that give rise to mathematically formulated criteria which the separate processes or the theoretical representations of them have to satisfy" [4, p. 228]. Einstein's emphasis on "mathematically formulated criteria" is crucial, and it played a major role in Heisenberg's and Dirac's thinking as well. Thermodynamics, Einstein's example of a classical principle theory (parallel to the kinetic theory of gases as a constructive theory), is a principle theory because it "seeks by analytical means to deduce necessary conditions, which separate events have to satisfy, from the universally experienced fact that perpetual motion is impossible" [4, p. 228]. Heisenberg's thinking leading to his discovery of quantum mechanics was similarly principle, and perhaps influenced by thermodynamics, along with (this is known) Einstein's special relativity, which was a principle theory as well. Einstein, it may be added, implied the kinetic theory of gases could be not be derived from thermodynamics. He then adopted a parallel view that a proper, constructive theory of quantum processes (which he wanted to be a theory of continuous classical-like fields) could not be derived from quantum mechanics.

Principles, then, are "empirically discovered, general characteristics of natural processes, … that give rise to mathematically formulated criteria which the separate processes or the theoretical representations of them have to satisfy." I shall adopt this definition of "principle" here, including the requirement of "mathematically formulated criteria" to which empirically discovered elements and physical principles give rise, a requirement that is crucial and plays a major role in the present analysis. I shall, however, add the following qualification (which would probably have been accepted by Einstein). Principles are not so much empirically discovered as formulated, *constructed*, on the basis of empirically discovered or established features of scientific experience. (This "construction," as is that of the mathematics of a principle theory, is of course different from the synthetic theoretical construction of an underlying theoretical scheme on the basis of which the phenomena considered might be explained, in a constructive theory.) As Einstein argued on many occasions (against empiricism, such as that of Mach), it would be difficult to see "general characteristics of natural processes" as ever merely empirically given. "The impossibility of perpetual motion" could hardly be seen as empirically given; it was instead formulated, as a principle, on the basis of empirically established evidence. Principles, thus, need not have the self-evidence of axioms or, at least initially, the assumption-like character of postulate, although, once introduced, they may function as postulates or even axioms from which a given theory is built by means of logical rules and deductions. This article will further amplify this understanding of principles, by seeing a principle, as it was by Bohr, Heisenberg, and Dirac, as a foundation and guidance for in inventing and building a new theory.

Einstein, who by the time of quantum mechanics had developed a strong preference for constructive theories, hoped that quantum mechanics, as a principle theory, would be replaced by a constructive theory, closer to E. Schrödinger's initial wave approach (which is constructive), much preferred by Einstein, vs. Heisenberg's matrix version (which is principle). In addition, Schrödinger's wave mechanics was very much in accord with Einstein's own program, fully in place by then, for a unified field theory, which aimed to derive quantum discreteness from a continuous classical-like field theory (which Einstein was never able to do, any more than Schrödinger was, in his several attempts at such a theory). By contrast, as will be discussed in detail below, for Heisenberg, quantum discreteness was one of the primary postulates and principle, although the concept requires a complex interpretation, ultimately developed by Bohr. While less committed to the constructive approach at the time than he eventually became, Einstein entertained a similar hope in his initial work on the old quantum theory, which was not constructive or even principle, as his special relativity was, but rather, as he called it then, "heuristic" [14]. For a while Einstein saw his theoretical thinking, intriguingly along with that of Bohr, as primarily principle, and even at the time of his introduction of this distinction in 1919, Einstein still recognized the pretty much equal value of each type of theory. However, his preference for constructive theories, defined by "free conceptual construction" and mathematics that embodies this construction (rather than only giving mathematical expression to physical principles) grew, ultimately to the point of a near unconditional insistence on them, which defined Einstein's thinking by the time quantum mechanics entered the picture [15, p. 47; 16].

Constructive theories tend to be, and are often aimed to be, realist or ontological, insofar as they describe, usually causally, the corresponding objects in nature and their behavior by way of mathematical models, assumed to idealize how nature works at the simpler, or deeper, level thus constructed by a theory. This characterization will serve in this article as the definition of a realist or ontological theory. Such a theory, thus, offers a description of the processes underlying and connecting the observable phenomena considered, on the model of classical mechanics, from which quantum theory departs, thereby also divorcing the theory from the *description* of the observed phenomena and connections between them, and hence relating to quantum phenomena and the reality underlying them otherwise. By "reality" itself I shall refer to that which actually exists or is assumed to exist. In the case of physics, it is, ultimately, nature or matter, which is generally, but not always, assumed to exist independently of our interaction with it, and to have existed when we did not exist and to continue to exist when we will no longer exist. What is that which exists or is assumed to exist, and how it exists, is a matter of perspective, interpretation, and debate.

According to this definition, realism or ontology is not only a claim concerning the existence of something but also and primarily a claim concerning the character of this existence. The definitions of reality, realism, and ontology just given cannot claim to capture all of the deeper physical and philosophical aspects of these concepts. They are, however, sufficient for my purposes. I might add that all modern, post-Galilean, physical theories proceed by way of idealized mathematical models, even if these models are not realist, descriptive. Some of the models used in quantum theory or its interpretations are of this nonrealist kind, insofar as, according to them, quantum mechanics or higher-level quantum theories are only predictive, and moreover only probabilistically predictive. Both Heisenberg and Dirac adopt this latter view of quantum mechanics or quantum electrodynamics, following Bohr and "the spirit of Copenhagen," as Heisenberg called it [2, p. iv].[3] This suggests a general definition of a mathematical model, which I shall adopt here. A mathematical model is a mathematical structure or set of mathematical structures that enables any type of relation, descriptive or predictive, to the (observed) phenomena or objects (which need not be observable, and quantum objects are not) considered. Such a model is always *constructed*, although it may be that of either a constructive or a principle *physical* theory. Realist mathematical models are descriptive models—idealized mathematical descriptions of physical processes. It may be noted that the probabilistic character of quantum predictions must be equally maintained by realist interpretations of these theories or alternative theories (such as Bohmian theories), because it corresponds to what is actually observed in quantum experiments, concerning which only probabilistic or statistical predictions are possible. This is because the repetition of identically prepared experiments, in general, leads to different outcomes, a difference that, unlike in classical physics, cannot be improved beyond a certain limit (defined by Planck's constant, $h$) by improving the conditions of measurement, a fact also reflected in the uncertainty relations. This constitutes the quantum probability principle, the QP principle, used by Heisenberg in his discovery of quantum mechanics and given by him a mathematical expression, explained below. On the other hand, even such predictive interpretations generally assume, as those of Bohr, Heisenberg, and Dirac did, the concept of reality. If, however, realism, as just defined, presupposes a description or at least a conception of reality, this concept of reality is that of "reality without realism," the concept and principle, the RWR principle, that implies that a theory that follows this principle is a principle theory, although it may also be defined by other principles.[4]

---

[3] I distinguish "the spirit of Copenhagen" from "the Copenhagen interpretation," a rubric that I shall avoid, because there is no single such interpretation. Indeed, some interpretations designated "Copenhagen interpretations" only partially conform to the spirit of Copenhagen as understood here, and some do not conform to it at all.
[4] For a further discussion of the concept of reality without realism, see [17].

Einstein sees his special relativity theory as a principle theory, based in two apparently, but only apparently, irreconcilable principles. The first is the principle of relativity, which says, in Einstein's initial formulation, that "the same laws of electrodynamics and optics will be valid for all frames of reference for which the equation of mechanics hold good" and the second is the principle of the constancy of the velocity of light, regardless of the motion of the source [18, p. 37]. These two principles are only apparently irreconcilable: special relativity theory reconciles them, a reconciliation that was Einstein's great achievement. Einstein's general relativity, his non-Newtonian theory of gravity, is both a principle and constructive theory. It is a principle theory because it is based most essentially in the equivalence principle (postulating the equivalence of inertial and gravitational mass). However, it also has a constructive and realist dimension because it represents, *constructs*, the physical nature of gravity as the curvature of space or spacetime (curved by the presence of physical bodies or fields) and describes the behavior of its objects accordingly, in a realist and causal manner. Accordingly, the mathematical embodiment of the principle of general relativity, as well as a (realist) model of the theory, is Riemannian geometry of, in general, variable curvature. Indeed, while in the article on constructive and principle theories under discussion and elsewhere in his earlier commentaries on the theory Einstein emphasized its principle aspects, eventually its constructive aspects took the primary significance in Einstein's assessment of theory and in his thinking in general, as considered in [16].

It follows that a principle theory could be either realist or not, in the first case unavoidably bringing with it a constructive dimension, unless the phenomena or objects in question are already given, rather than being constructed as the simpler constituents of more complex phenomena, which is to say, have already been constructed. Constructive theories are, as I explained, nearly by definition realist and are usually causal, unless one uses a given construction as a kind of heuristic device within a predictive (principle) theory. It is also true that a given theoretical construction may be revealed or argued merely to provide a predictive mechanism for a given theory, and there are arguments to that effect concerning the status of spacetimes of general relativity [19]. That, however, is not the same as developing a given theory as a constructive one.

Constructive theories may and often do involve principles, such as the equivalence principle in general relativity, or the principle of causality, found throughout modern physics from Galileo on until quantum mechanics put it into question. This principle, as defined, for example, by Kant (this definition has been commonly used since), states that, if an event takes place, it has a cause of which it is an effect [20, pp. 305, 308]. The principle of causality is operative in most constructive theories at two levels. First, the construction itself is the application, in this case more philosophical, of the principle of causality, insofar as "building up a picture of the more complex phenomena out of the materials of a relatively simple formal scheme from which they start out," presupposes a cause-effect relationship. Secondly, and this application of the principle of causality is physical, the formal scheme in question is usually that of an (idealized) causal process, in which the state of the system considered at a given moment of time determines its state at all future moments of time. I shall refer to this form of causality as *classical* causality.[5]

---

[5] I distinguish causality, which is an ontological category, describing reality, from determinism, which is an epistemological category, describing part of our knowledge of reality, specifically our ability to predict the state of a system, at least as defined by an idealized model, exactly at any moment of time once we know its state at a given moment of time. Determinism is sometimes used in the same sense as causality, and in the case of classical mechanics (which deals with single objects or a sufficiently small number of objects), causality and determinism, as

Such causal influences are also commonly, although not always, assumed to propagate from past or present towards future. This requirement is strengthened by special relativity theory, which further restricts causes to those occurring in the backward (past) light cone of the event that is seen as an effect of this cause, while no event can be a cause of any event outside the forward (future) light cone of that event. These restrictions follow from the assumption that causal influences cannot travel faster than the speed of light in a vacuum, *c*. Principle theories do not require classical causality, which indeed becomes difficult, if not impossible, to assume in quantum physics. Relativistic "causality," as the prohibition of the possibility of physical influences towards the past is usually maintained, although there are exceptions (e.g., [21, pp. 182-209], and it may even be adopted as or linked with a principle of causality in quantum theory, without presuming classical causality.[6] It follows that the distinction between constructive and principle theories is not unconditional, as Einstein came to realize as well. This led him to ever more complex schemes of theoretical practice in fundamental physics, while ultimately preferring a constructive approach, grounded, moreover, in the idea (the principle?) that this construction and the corresponding physical reality should emerge from a free invention of the mathematical scheme with only minimal, if any, connections to the experiment, as discussed in detail in [16]. There is, however, an asymmetry between these two types of theories: a constructive theory always involves principles, at least philosophical principles, while a principle theory need not involve constructive strata at the ultimate level considered by the theory, as quantum mechanics came to show, arguably, for the first time.

Quantum mechanics or higher-level quantum theories are principle theories, at least if one follows the spirit of Copenhagen in interpreting them, again, the first theory that, in this interpretation, is strictly, irreducibly principle insofar as it precludes the claim for the constructive theorization of quantum objects and processes. Although, quantum mechanics has rarely been expressly described as a principle theory on Einstein's definition just given (e.g., [6]), this nature of quantum mechanics, at least matrix mechanics, in interpretations in the spirit of Copenhagen was manifested from the outset, in contrast to Schrödinger's wave mechanics. As noted above, the latter was constructive, even though it proceeded from principles as well. Schrödinger's mathematics, equivalent to that of Heisenberg, could be and, beginning with Bohr and Heisenberg, has been interpreted in the spirit of Copenhagen. Indeed, if understood in this spirit, quantum mechanics is a principle theory by definition, because of the fundamentally nonrealist character of the corresponding interpretations, such as that of Bohr, or more accurately (because Bohr's views evolved), Bohr's ultimate interpretation. In this type of interpretation, it is not possible to "construct" the ultimate entities, quantum objects, from which the observable quantum phenomena are built or due to which these phenomena arise, unless one sees quantum objects as *constructed* by quantum theory as fundamentally *unconstructible*. It follows that, in this type of interpretation, quantum mechanics or higher-level quantum theories are nonrealist, and that in the corresponding interpretation of quantum phenomena themselves, a realist or, by the same token, constructive understanding of the ultimate ("quantum") constitution of nature responsible for this phenomena is, in Bohr's words, "*in principle* excluded" [24, v. 2., p. 62].

---

defined here, coincide. Once a system is large enough, one needs a superhuman power to predict its behavior exactly, as was famously noted by P. S. Laplace. However, while it follows automatically that noncausal behavior, *considered at the level of a given model*, cannot be handled deterministically, the reverse is not true. The underlined qualification is necessary because we can have causal models of processes in nature that may not be causal.

[6] As will be seen, D'Ariano et al adopt a principle of causality of that type [10, p. 3, 11]. See also [22] and [23] for further discussions of causality in quantum theory.

Thus, quantum theories divorce quantum theory from the *description* of observed phenomena and connections, overt (as in classical mechanics) or hidden, between them, and relate to quantum phenomena otherwise, specifically in terms of predictions, which, it follows, become, in general, probabilistic or statistical in nature.[7] Einstein hoped and indeed assumed that nature will eventually allows us to do better as concerns the descriptive or realist capacity of quantum theory or, as Einstein thought, an alternative theory that would eventually replace it. Bohr, on the other hand, thought that nature *might not* allow us to do better in dealing with quantum phenomena, which is not the same as that it *will not*, but Bohr, unlike Einstein, was not worried by this lack of realism or causality. These two positions define the Bohr-Einstein debate, by the time which (it began around 1927), Einstein's commitment to the constructive approach vs. the principle one and to the primary role of mathematics in fundamental theories was nearly absolute. Bohr, by contrast, has always remained committed to physical principles and to experimental evidence as the primary measures of fundamental theories, although their mathematical character was essential for him as well. The question whether nature might or might not allow for a constructive alternative to quantum mechanics (considered as a strictly principle theory) remains open and continues to be intensely and even fiercely debated. Most, however, hold on to Einstein's hope, and the type of view adopted by Bohr or here remains a minority view.

Bohr's phrase "*in principle* excluded," could and, I would argue, should be read not only as expressing a strong prohibition on the ultimate reach of our knowledge and even thought concerning quantum objects and their behavior, but also as an annunciation of a new principle of quantum reality, defined above as the principle of "reality without realism," the RWR principle. This principle maintains both the existence, *reality*, of quantum objects (or certain entities in nature that are thus idealized) and the impossibility of representing or even conceiving of the nature of this reality, and hence the impossibility of realism, at least as things stand now (a crucial qualification assumed throughout this article).[8] As I shall explain in more detail below, as a consequence of the RWR principle, classical causality becomes automatically impossible and the recourse to probability, accordingly unavoidable even in considering the individual quantum processes and events, in conformity with the QP principle. However, this principle, even though it can, as will be seen, be given a (complex) mathematical expression, is an interpretive rather than empirical principle, and historically, it emerged when quantum mechanics was already in place. Heisenberg, in his initial approach to quantum mechanics, merely abandoned the project of describing the motion of electrons themselves because he had seen such a description as unachievable at the time, rather than had assumed that such a description was or might be "*in principle* excluded" [25]. His approach may, however, be seen as guided by the combination of a certain "proto-RWR" principle and the QP principle, as well as the principle of quantum discreteness (QD). As noted earlier, Schrödinger, in developing his wave mechanics, aimed at a constructive and realist theory. The RWR principle emerged, sometime in the mid 1930s in the Bohr-Einstein debate, especially in the wake of their exchange concerning the Einstein-Podolsky-Rosen (EPR) experiment, introduced in 1935, which also related the question of reality to that of locality [26, 27].[9] This exchange, which occurred at the height of Einstein's work on a (classical-like) unified field theory, reflects and is defined by their contrasting commitments to the constructive (Einstein) and the principle (Bohr) approach to fundamental theories in physics.

---

[7] In doing so, quantum theory also suggested that it may true more generally, for example, as noted earlier, in general relativity [19], but the subject would require a separate treatment.

[8] Although Bohr does not appear to have made or possibly subscribed to the stronger claim, assumed in this article, of the impossibility of even any conception of quantum objects and their behavior, this claim may be seen as made in the spirit of Copenhagen. Not all interpretations in the spirit of Copenhagen adopt this more radical view and some stop short of that of Bohr.

An appeal to fundamental principles need not imply that there is some unchanging, Platonist-like, essence to such principles. Principles change as our experimental findings and our theories change, and we cannot anticipate or control all of these changes. Indeed, while having confidence in a given set of principles can help and drive one's creative work, an uncritical, dogmatic acceptance of any such set can inhibit and even prevent an advancement of thought and knowledge, which may require new principles, the invention which is, I argue, one of the greatest forms of creativity in physics and beyond. The principles of quantum mechanics, such as the QP principle, replaced, within a new scope, some among the main principles of classical physics, which continue to remain operative within the proper scope of classical physics. Some of the principles of classical physics, however, extend to quantum mechanics, such as the principle that a physical theory must be mathematical, which defined all modern physics, classical, relativistic, or quantum. There could be such changes within the same physical scope, as in the case of general relativity theory vs. Newton's theory of gravity, based in classical mechanics. Some of the principles of quantum theory have changed as well, although here the development of the theory after quantum mechanics (which was a change from the old quantum theory within the same scope) was only defined by expanding the scope of quantum theory. The QP principle has remained in place throughout the history of quantum mechanics, from Heisenberg on, although the mathematical expression of the principle was refined a few times, which fact tells us that there is no Platonist-like essence to the connections between mathematics and physics either, or even to mathematics itself, which changes, too. On the other hand, the correspondence principle has changed in its definition and functioning, by becoming the "mathematical correspondence principle," in Heisenberg's initial work on quantum mechanics, and it became even more general in Dirac's work. While it was intimated by Heisenberg in his paper introducing quantum mechanics and while it could have been inferred from the complementarity principle, introduced in 1927, the RWR principle emerged a decade later in Bohr's work, as part of his, by then changed in turn, interpretation of quantum mechanics. Later on, sometime in the1990s, the principles of quantum theory were developed into those of quantum information theory, to be discussed in Section 5. Combining the principles of relativity and quantum mechanics by giving precedence to the principles of quantum mechanics (as against the Klein-Gordon theory) led Dirac to the discovery of his equation. The discovery of antimatter, which was an unexpected bonus, eventually led to yet new principles, grounding quantum field theory and the Standard Model of elementary particle physics.

### 3. Heisenberg, Bohr, and the Principles of Quantum Mechanics

In both Heisenberg's initial approach to quantum mechanics and Bohr's initial interpretation of the theory the key principles were:

(1) the principle of quantum discreteness or the QD principle (initially defined by what Bohr called "the quantum postulate," "symbolized," by Planck's quantum of action, $h$ [24, v. 1, p. 53-

---

[9] Bell's and the Kochen-Specker theorems, dealing with the EPR-type phenomena (for discrete variables) may be seen as lending support to this principle, in part in view of the question of locality, insofar as the latter could be maintained if one adopt the RWR-principle. The meaning and implications of these theorems and related findings, and of the concepts involved, such as locality, are under debate. I shall return to the question of locality in Section 5.

54]), according to which all observed quantum phenomena are individual and discrete (in relation to each other), which is not the same as the (Democretian) atomic discreteness of quantum objects themselves;[10]

(2) the principle of the probabilistic or statistical nature of quantum predictions, even, in contradistinction to classical statistical mechanics, in the case of elemental, unsubdivisible, quantum processes and events—the quantum probability, QP, principle, and

(3) the correspondence principle, which, as initially used by Bohr required that the predictions of quantum theory must coincide with those of classical mechanics at the classical limit (even though the processes themselves were still quantum at this limit), but which was given by Heisenberg a new, more rigorous and more precise form, defined here as "the mathematical correspondence principle," requiring that the equations of quantum mechanics convert into those of classical mechanics at the classical limit.[11]

The correspondence principle was not a quantum principle, because it did not reflect the quantum constitution of nature, but it played an essential role in Heisenberg's development of the mathematical formalism of (matrix) quantum mechanics.

Bohr's interpretation added a new principle:

(4) the complementarity principle, which stems from the concept of complementarity, introduced by Bohr a bit later (following Heisenberg's discovery of the uncertainty relations), and which requires: *(a)* a mutual exclusivity of certain phenomena, entities, or conceptions; and yet *(b)* the possibility of applying each one of them separately at any given point, and *(c)* the necessity of using all of them at different moments for a comprehensive account of the totality of phenomena that one must consider in quantum physics.

Principles 1 and 2 are correlative, although this was understood only in retrospect, especially with the development of Bohr's interpretation, eventually leading to the RWR principle, via the uncertainty relations and the complementarity principle.[12] The RWR principle could be inferred from the complementarity principle, because the latter prevents us from ascertaining the complete composition of the "whole from parts," to the degree this concept applies, because the complementary parts never add to a whole in the way they do in classical physics or relativity. This became especially apparent in view of the EPR-type experiments, a point to which I shall return in Section 5. The uncertainty relations are sometimes seen as manifesting a principle as well, but this view requires significant qualifications and will not be adopted here.

While Heisenberg did not expressly refer to either the QD or the QP principle as such, both principles were manifestly at work in his derivation of quantum mechanics. The QP principle, which can be given a mathematical expression, could be considered as a primitive principle, primary to the RWR principle or, arguably, to any other quantum-theoretical principle. The RWR principle is, as I said, an interpretive inference from the QP principle, which would apply even in the case of causal theories of quantum phenomena, such as Bohmian mechanics. The latter, a constructive theory (in all of its versions), does not conform to the RWR principle and entails both realism and causality at the quantum level, even though it retains, correlatively, both the uncertainty relations and the irreducibly probabilistic character of quantum predictions, which, in Bohmian mechanics, strictly coincide with those of quantum mechanics. In classical physics, when the recourse to probability is involved, we proceed from causality to probability because of our inability to track this causality. This, while leaving room for probability, gives primacy to the principle of causality, which makes probability reducible, at least ideally and in principle, specifically in considering individual classical events. In quantum mechanics, in Heisenberg's and most other derivations, probability is, again, a primitive concept and the QP principle is a primitive principle, as is, correlatively, the QD principle, while the RWR principle is interpretively inferred from the QP principle, although one could assume it to be a grounding principle, and Heisenberg, again, nearly did. The absence of classical causality is an automatic consequence of the RWR principle. As Schrödinger observed, by this point (in 1935) with some disparagement, if there is no definable physical state, and there is none by the RWR principle, one cannot assume that it changes (classically) causally [30, p. 154]. There is only relativistic causality, which does not require classical causality, although relativity itself is a classically causal theory.

---

[10] In the late 1930s, following his exchanges with Einstein concerning the EPR-type experiments, Bohr rethought this principle in terms of his concept of phenomenon [24, v. 2, p. 64]. For the discussion of Bohr's views by the present author, see [28] and [29].

[11] I have considered Heisenberg's discovery of quantum mechanics, in terms of principles, in detail in [7, pp. 77-137]. See also [6] for a related but a somewhat different view of Bohr's key principles.

[12] The principle of complementarity, as formulated here, reflects more Bohr's later works, from 1929 on (the concept itself was introduced in 1927), impacted by his debate with Einstein. In these works the principle is exemplified by the complementary nature of the position and the momentum measurements, always mutually exclusive and as such correlative to the uncertainty relations. See [28, 29], for the development of Bohr's views.

I shall now consider how the principles just described worked and were given a mathematical expression by Heisenberg in his discovery of (matrix) quantum mechanics, as presented in his original paper [25].[13] Admittedly, this mathematical expression was only partially worked out and sometimes more intuited than properly developed. Nevertheless, Heisenberg's creativity and inventiveness were remarkable. Bohr's initial comment on Heisenberg's discovery, in 1925 (before Schrödinger's version was introduced), shows a clear grasp of what was at stake: "In contrast to ordinary mechanics, the new quantum mechanics does not deal with a space–time description of the motion of atomic particles. It operates with manifolds of quantities which replace the harmonic oscillating components of the motion and symbolize the possibilities of transitions between stationary states in conformity with the correspondence principle. These quantities satisfy certain relations which take the place of the mechanical equations of motion and the quantization rules [of the old quantum theory]" [24, v. 1, p. 48].

In reflecting on Heisenberg's conception of these quantities, one might observe first that, in order to invent a new concept of any kind, one has to construct a phenomenological entity or a set of such entities and relations between them. In physics, one must also give this construction a mathematical architecture, with which one might indeed start, as Heisenberg in effect did, and which enables the theory to *relate* to observable phenomena and measurable quantities associated with these phenomena. Heisenberg's approach, as that of Dirac later on (influenced by Heisenberg), may even be best seen as defined by an attempt to find, first, under the guidance of the principles assumed, an independent (abstract) mathematical scheme that would then be linked, in his case (or, again, that of Dirac) to the data obtained in the quantum experiments in question. This approach as such is general, rather than specifically quantum-mechanical, in nature, and it was in fact close to that of Einstein's more mathematical thinking, still guided by the equivalence principle (which is physical), in the case of general relativity, in partial contrast to special relativity, where mathematics was less dominant. However, again, contrary to Einstein's desideratum, a theory and its phenomenal-mathematical architecture need not describe, in the way it does or at least may be assumed to do in classical physics or relativity, observable physical phenomena and connections between them (hence my emphasis on "relate"). In quantum mechanics, or quantum field theory, this architecture only relates to the observed quantum phenomena probabilistically or statistically, which nevertheless allows the theory to remain a mathematical-experimental science of nature, in conformity with the project of modern physics from Galileo on.

---

[13] For details, see [7 pp. 77-137].

Heisenberg's invention of his matrices was enabled by his idea of arranging algebraic elements corresponding to numerical quantities (transition probabilities) into infinite square tables. It is true that, once one deals with the transitions between two stationary states, rather than with a description of such states, matrices appear naturally, with rows and columns linked to each possible state respectively. This naturalness, however, became apparent or one might say, *became natural*, only in retrospect. This arrangement was a phenomenological construction, which amounted to that of a mathematical object, a matrix, an element of general noncommutative mathematical structure, part of (infinite-dimensional) linear algebra, in effect entailing a tensor structure of a Hilbert space, in which Heisenberg's observables form an operator algebra. One can also see it as a representation of an abstract algebra, keeping in mind that Heisenberg's infinite matrices were unbounded, which fact, as became apparent shortly thereafter, is correlativity the uncertainty relations for the corresponding continuous variables.

Heisenberg begins his derivation with an observation along the lines of a proto-RWR principle: "[I]n quantum theory it has not been possible to associate the electron with a point in space, *considered as a function of time*, by means of observable quantities. However, even in quantum theory it is possible to ascribe to an electron the emission of radiation" [25, p. 263; emphasis added]. I add emphasis because, in principle, a measurement could associate an electron with a point in space, but, which is the main point here, not as a function of time, in the way it can be done in classical mechanics. Note also that we cannot definitively establish the moment of a given emission of radiation but can only register the outcome of this emission at a later time. If one adopts Bohr's interpretation in its ultimate form, in which one cannot assign any properties to quantum objects themselves (not even single such properties, rather than only certain joint ones, which is precluded by the uncertainty relations) but only to the measuring instruments involved, one can see Heisenberg's discovery in quantum-informational terms. It amounted to establishing a mathematical scheme that enables the processing of information (which is, qua information, classical) between measuring devices. I shall return to this view of the quantum-mechanical situation later. Heisenberg then says: "In order to characterize this radiation we first need the frequencies which appear as functions of two variables. In quantum theory these functions are in the form [originally introduced by Bohr]:

$$v(n, n - \alpha) = 1/h \, \{W(n) - W(n - \alpha)\} \quad (1)$$

and in classical theory in the form

$$v(n, \alpha) = \alpha v(n) = \alpha/h(dW/dn)" \text{ [25, p. 263].}$$

This difference, reflecting what I called above the quantum discreteness (QD) principle, introduced by Bohr and understood by him in terms of the quantum postulate following the introduction of quantum mechanics [24, v. 1, p. 53], leads to a difference between classical and quantum theories as concerns the combination relations for frequencies, which correspond to the Rydberg-Ritz combination rules. However, "in order to complete the description of radiation [in accordance, by the mathematical correspondence principle, with the Fourier representation of classical kinematics] it is necessary to have not only frequencies but also the amplitudes" [25, p.

263]. The crucial point is that, in Heisenberg's theory and in quantum mechanics since then, these "amplitudes" are no longer amplitudes of any physical motions, which makes the name "amplitude" itself an artificial, *symbolic* term. These amplitudes are linked to the probabilities of transitions between stationary states: they are what we now call probability amplitudes. The corresponding probabilities are derived, from Heisenberg's matrices, by a form of Born's rule for this limited case (Born's rule is more general). One takes square moduli of the eigenvalues of these matrices (or equivalently, multiply these eigenvalues by their complex conjugates), which gives one real numbers, corresponding, once suitably normalized, to the probabilities of observed events. (Technically, one also needs the probability density functions, but this does not affect the essential point in question.) The standard rule for adding the probabilities of alternative outcomes is changed to adding the corresponding amplitudes and deriving the final probability by squaring the modulus of the sum. The mathematical structure thus emerging is in effect that of vectors and (in general, noncommuting) Hermitian operators in Hilbert spaces over complex numbers, which spaces are in this case, infinite-dimensional, given that we deal with continuous variable. This structure may be seen as a mathematical expression of the QP principle. Heisenberg explains the situation in these, more rigorous, terms in his 1930 book [2, pp.111-122]. In his original paper, he argues as follows:

> The amplitudes may be treated as complex vectors, each determined by six independent components, and they determine both the polarization and the phase. As the amplitudes are also functions of the two variables $n$ and $\alpha$, the corresponding part of the radiation is given by the following expressions:

Quantum-theoretical:

$$\mathrm{Re}\{A(n, n - \alpha)\mathrm{e}^{i_\omega(n, n - \alpha)t}\}$$

Classical:

$$\mathrm{Re}\{A_\alpha(n)\mathrm{e}^{i_\omega(n)_\alpha t}\} \text{ [25, p. 263]}$$

The problem—a difficult and, "at first sight," even insurmountable problem—is now apparent: "[T]he phase contained in $A$ would seem to be devoid of physical significance in quantum theory, since in this theory frequencies are in general not commensurable with their harmonics" [25, pp. 263-264]. Heisenberg now proceeds to inventing a new theory around this problem, in effect, by making it into a solution, as if saying: "This is not a problem, the classical way of thinking is." His new theory offers the possibility of predicting, in general probabilistically, the outcomes of quantum experiments, but at the cost of abandoning the physical description of the ultimate objects considered, a cost unacceptable to some, even to most, beginning with Einstein, but a new principle for Bohr, the RWR principle. Heisenberg says: "However, we shall see presently that also in quantum theory the phase has a definitive significance which is *analogous* to its significance in quantum theory" [25, p. 264; emphasis added]. "Analogous" could only mean here that, rather than being analogous physically, the way the phase functions mathematically is analogous to the way the classical phase functions mathematically in classical theory, or analogous in accordance with the *mathematical* form of the correspondence principle, insofar as quantum-mechanical equations are formally the same as those of classical physics. Heisenberg only considered a toy model of a quantum aharmonic oscillator, and thus needed only a Newtonian, rather than Hamiltonian, equation for it.

In this way, Heisenberg gave the correspondence principle a mathematical expression, indeed changed it into the mathematical correspondence principle. The variables to which these equations apply cannot, however, be the same, because, if they were, they would not give us correct predictions for low quantum numbers. As Heisenberg explains, if one considers "a given quantity $x(t)$ [a coordinate as a function of time] in classical theory, this can be regarded as represented by a set of quantities of the form

$$A_\alpha(n)e^{i_\omega(n)_\alpha t},$$

which, depending upon whether the motion is periodic or not, can be combined into a sum or integral which represents $x(t)$:

$$x(n,\ t) = \sum_\alpha^{+\infty}{}_{-\infty} A_\alpha(n)\ e^{i_\omega(n)_\alpha t}$$

or

$$x(n,\ t) = \int_{-\infty}^{+\infty} A_\alpha(n)\ e^{i_\omega(n)_\alpha t}\ d\alpha\ "\quad [25,\ p.\ 264].$$

Heisenberg next makes his most decisive and most extraordinary move. He notes that "a similar combination of the corresponding quantum-theoretical quantities seems to be impossible in a unique manner and therefore not meaningful, in view of the equal weight of the variables $n$ and $n - \alpha$" (25, p. 264). "However," he says, "one might readily regard the ensemble of quantities $A(n, n - \alpha)e^{i_\omega(n,\ n\ -\ \alpha)t}$ [an infinite square matrix] as a representation of the quantity $x(t)$" [25, p. 264].

The arrangement of the data into square tables is a brilliant and, as I said, in retrospect, but, again, only in retrospect, natural way to connect the relationships (transitions) between two stationary states. However, it does not by itself establish an *algebra* of these arrangements, for which one needs to find the rigorous rules for adding and multiplying these elements—rules without which Heisenberg cannot use his new variables in the equations of the new mechanics. To produce a *quantum-theoretical interpretation* (which, again, abandons motion and other concepts of classical physics at the quantum level) of the classical equation of motion that he considered, as applied to these new variables, Heisenberg needs to be able to construct the powers of such quantities, beginning with $x(t)^2$, which is actually all that he needs for his equation. The answer in classical theory is obvious and, for the reasons just explained, obviously unworkable in quantum theory. Now, "in quantum theory," Heisenberg proposes, "it seems that the simplest and most natural assumption would be to replace classical [Fourier] equations … by

$$B(n,\ n-\beta)e^{i_\omega(n,\ n\ -\ \beta)t} = \sum_\alpha^{+\infty}{}_{-\infty} A(n,\ n-\alpha)A(n-\alpha,\ n-\beta)e^{i_\omega(n,\ n\ -\ \beta)t}$$

or

$$= \int_{-\infty}^{+\infty} A(n,\ n-\alpha)A(n-\alpha,\ n-\beta)e^{i_\omega(n,\ n\ -\ \beta)t}\ d\alpha"\ [25,\ p.\ 265].$$

This is the main postulate, the (matrix) multiplication postulate, of Heisenberg's new theory, "and in fact this type of combination is an almost necessary consequence of the frequency combination rules" [25, p. 265]. This combination of the particular arrangement of the data and the construction of an algebra of multiplying his new variables is Heisenberg's great invention. Although, it is commutative in the case of squaring a given variable, $x^2$, this multiplication is in general noncommutative, expressly for position and momentum variables, and Heisenberg, without quite realizing it, used this noncommutativity in solving his equation, as Dirac was the first to notice. Taking his inspiration from Einstein's "new kinematics" of special relativity, Heisenberg spoke of his new algebra of matrices as the "new kinematics." This was not the best choice of term because his new variables no longer described or were even related to motion as the term kinematic would suggest, one of many, historically understandable, but potentially confusing terms. (Note that Planck's constant, $h$, which is a dimensional, dynamic entity, has played no role thus far.) Technically, the theory, as Einstein never stopped noting, wasn't even a mechanics, insofar it did not offer a description of individual quantum processes, or for that matter of anything. "Observables," for the corresponding operators, and "states," for Hilbert-space vectors, are other such terms: we never observe these "observables" or "states," but only use them to predict, probabilistically, what is observed in measuring instruments. (To make these predictions, one will need $h$, which, we recall, appears in Bohr's frequency rules.)

As noted above, Heisenberg's overall scheme essentially amounts to the Hilbert-space formalism (with Heisenberg's matrices as operators), introduced by J. von Neumann shortly thereafter, thus giving firm and rigorous mathematical foundations to Heisenberg's scheme, by then developed more properly by Heisenberg himself, M. Born and P. Jordan, and, differently (in terms of $q$-numbers*), by Dirac. My main point is that Heisenberg's matrices were (re)invented by him from the physical principles coupled to a mathematical construction leading to an actual algebra, which Heisenberg had to define, beginning with the multiplication rule. This multiplication is noncommutative and ultimately implies tensor calculus in a Hilbert space, as Heisenberg was to learn later on [2, pp. 55-56, 111-122]. Dirac, who followed Heisenberg's principle way of thinking in his work on both quantum mechanics and quantum electrodynamics, was also the first to fully realize that noncommutativity was the most essential mathematical feature of Heisenberg's scheme. Remarkably, Heisenberg himself, as well as Pauli, not only did not think it to be essential but also thought that ultimately the theory should be freed from it, and Pauli initially thought that the theory should not be probabilistic either, and changed his mind on both counts only after Schrödinger's equation was introduced [7, pp. 89-90].

The physical principle behind quantum noncommutativity and, by implication, the tensor structure of quantum theory is a more complex matter. In fact, it is Bohr's complementarity principle. Conversely, quantum noncommutativity and this tensor structure can be seen as the mathematical expression of the complementarity principle, even though noncommutativity was discovered first, in part as a response to the QP principle. As noted above, the QP principle itself is given its mathematical expression, via the complex Hilbert-space structure cum conjugation, inherent in this structure, and Born's rule. This structure is, however, in turn coupled to complementarity, a coupling manifested in the uncertainty relations. Finally, insofar as the complementarity principle implies, at least if one follows the spirit of Copenhagen, the RWR principle, too, is mathematically expressed in noncommutativity or, again, the tensor structure of the formalism of quantum theory.

We can thus see how the physical principles assumed by Heisenberg found their mathematical expression in his matrix mechanics. It is true that the relationships between the complementarity principle and the formalism emerged later. Also, the complementarity principle and the RWR principle were not expressed in the form just sketched by Bohr himself, who, unlike Heisenberg, was less concerned with giving a rigorous mathematical expression to his principles, although Bohr did realize the relationships between them and the formalism of quantum theory. Indeed, as I said, even in Heisenberg's case, these physical principles were given a preliminary and somewhat tentative, rather than fully rigorous, mathematical expression. This type of rigor would have been difficult before von Neumann gave the formalism its rigorous Hilbert-space form [31], although both Born and, especially, Dirac ($q$-numbers) made important steps in this direction, and Dirac's 1930 book nearly arrived there [3]. On the other hand, as von Neumann's title suggests (in contradistinction from those of Heisenberg's and Dirac's books [1, 2]), in the book are primarily "the *mathematical foundations* of quantum mechanics" rather than fundamental *physical principles* of quantum mechanics, even though these principles must, again, be given their mathematical expression. This is perhaps why von Neumann's work never especially appealed to Bohr. It would not be accurate to say that von Neumann does not deal at all with such principles. He does. But his main focus is elsewhere, on recasting quantum mechanics into a rigorous mathematical form, rather than deriving it from new physical principles, in the way Heisenberg and Dirac do in the case of, respectively, quantum mechanics and quantum electrodynamics, or in the way quantum information theory aims to do. "Foundations" is a more general category and may refer to either physical or mathematical, or philosophical, foundations of a theory, in the sense of what essentially grounds it.

## 4. Dirac's Equation: How to Combine, and How Not to Combine, the Principles of Relativity and Quantum Theory

Dirac's thinking in his approach to the problem of the free relativistic electron was shaped by the following considerations. In order to equally respect, as such an equation must, the principles of relativity and quantum theory, certain specific mathematical features were required, two of them in particular. While, however, formally mathematical (Dirac followed Heisenberg insofar as his aim was the invention, first, of a consistent mathematical scheme by means of which one could predict the outcomes of relevant experiments), these features reflected and emerged from fundamental physical principles of (special) relativity theory and quantum theory, as embodied in quantum mechanics. This made Dirac's theory, represented in his equation, a principle theory rather than a constructive theory. It was not constructive because, as quantum mechanics, if understood, as it was by Dirac, in the spirit of Copenhagen, Dirac's equation did not describe the motion of electrons in space and time, but only predicted the probabilities of registered events, those of collisions between electrons and measuring instruments, in accordance with the QP principle. The mathematical features in question were as follows. The first feature, stemming from *the principles of special relativity theory*, was that time and space must enter symmetrically and indeed that space and time must be interchangeable, which was not the case in Schrödinger's nonrelativistic equation for the electron, because it contained the first derivative of time and the second derivatives of coordinates. The second, stemming from *the principles of quantum theory* was that the equation must be a *first-order* linear differential equation in time, just as Schrödinger's equation was.

This feature is linked to several other key features of quantum-mechanical formalism, related to physical phenomena and the principles of quantum theory. Among these additional features were the noncommutativity of quantum variables, linear superposition or linearity in general, and the conservation of the probability current, which entails a positive definite probability density and which, combined with the first order derivative in time, may be seen as unitarity, following [11]. The QP principle is given a mathematical expression in Dirac's theory *analogously* to the way it was by Heisenberg in quantum mechanics, as discussed above, but within a more complex mathematical formalism employed by Dirac. This broader accord with quantum mechanics made Dirac's theory and his equation fundamentally quantum in character.

Another key affinity with Heisenberg's approach was Dirac's use of the mathematical correspondence principle (not, as I said, a quantum principle, but instrumental in building the mathematics of quantum mechanic), which was defined by the fact that at the classical limit the equations of quantum mechanics convert into those of classical physics. Dirac used the mathematical correspondence principle in his earlier work on quantum mechanics, inspired by Heisenberg's paper. Indeed, while Heisenberg did use the principle, it was Dirac who appears to have been the fist to expressly formulate it as the mathematical correspondence principle. In his first paper on quantum mechanics, he said: "The correspondence between the quantum and classical theories lies not so much in the limiting agreement when $h = 0$ as in the fact that the mathematical operations on the two theories obey in many cases the same [formal] laws" [32, p. 315]. Applying the principle in the case of his equation meant that at the nonrelativistic quantum limit this equation would convert into Schrödinger's equation. Proving this fact was mathematically more difficult than in Heisenberg's case, where the application of the principle was nearly automatic because he started with classical equations. Dirac's use of the principle also suggests the following extension of it. The mathematical formalism of a given higher-level quantum theory should at the lower limit convert into the mathematical formalism of the corresponding lower-level theory (usually already established). Thus, for example, in the case of string or brane theory, this means that its equations should, at the corresponding lower limit, covert into those of quantum field theory. As will be seen, D'Ariano and Perinotti's derivation of Dirac's equation might be seen as an enactment of this generalized mathematical correspondence principle, because Dirac's equation, which belongs to the usual (Fermi) scale of high-energy physics, "emerges from the large-scale [extending to Planck's scale] dynamics of the minimum-dimension QCA [quantum cellular automaton]" [11, p. 1]. That said, however, unlike Heisenberg (who borrowed his equation from classical physics, while inventing new quantum variables), Dirac did not use the mathematical correspondence principle, as opposed to other quantum principles just mentioned, to derive his equation, which was a new equation even formally, in this respect more akin to Schrödinger's equation.

Dirac's thinking leading his to his equation was further shaped by the following interrelated factors:

(1) *The influence of Heisenberg's thinking leading to the discovery of quantum mechanics, as discussed in Section 3*. Dirac studied Heisenberg's paper introducing quantum mechanics very carefully, and it left, I would argue, a lasting impact on all of his thinking and work. Heisenberg's paper was in many ways preliminary, and was developed into a full-fledged matrix mechanics a few months later by Heisenberg himself, M. Born, and P. Jordan. Dirac, unfamiliar with this subsequent work, arrived at his own, equally full-fledged, version, based in his $q$-number formalism, independently [27]. Heisenberg's paper, however, fully embodied these principles and introduced and refined some of them.

(2) *Dirac's work on his the transformation theory, his "darling," as he called it*. It was introduced by Dirac in 1926, while in Bohr's Institute in Copenhagen, and independently discovered by Jordan at the same time [33]. The transformation theory was especially important for Dirac's work on his equation, as concerned linearity in $\partial/\partial t$, and the positive definite probability density, both central to the transformation theory, which encoded the principles of quantum mechanics most generally, because it encompassed both Heisenberg's and Schrödinger's mechanics.

(3)   *Dirac's 1926 work on quantum electrodynamics (short of its relativistic form)* [34]. These factors made Dirac better prepared than others at the time for the discovery of his equation, which is not to say, quite the contrary, that the originality of his thinking had not played a decisive role in this discovery. Indeed, although Dirac's logic described above seems eminently reasonable in retrospect, it appears that only Dirac thought of the situation in this way at the time. His famous conversation with Bohr that occurred then is revealing:

Bohr: What are you working on?
Dirac: I am trying to get a relativistic theory of the electron.
Bohr: But Klein already solved that problem [35].

Dirac disagreed, and, for the reasons just explained, it is clear why he did, and why Bohr should have known better. The Klein-Gordon equation, to which Bohr referred, is relativistic and symmetrical in space and time, but it is not first-order linear differential equation in either, because both variables enter via the second derivative. One can derive the continuity equation from it, but the probability density is not positive definite. By the same token, the Klein-Gordon equation does not give us the correct equation, Schrödinger's equation

$$\frac{h}{2m}\nabla^2\psi = -ih\frac{\partial}{\partial t}\psi$$

at the nonrelativistic limit. Schrödinger, who appears to be the first to have written down the Klein-Gordon equation in the process of his discovery of his wave mechanics, abandoned it in view of the incorrect predictions it gave in the nonrelativistic limit. On the other hand, Dirac's equation, which could be seen as a square root of the Klein-Gordon equation, does convert into Schrödinger's equation in the nonrelativistic limit, which, again, was a major factor in Dirac's thinking. Technically, at its immediate nonrelativistic limit, Dirac's theory converts into Pauli's spin-matrix theory, while Schrödinger's equation, which does not contain spin, is the limit of Pauli's theory, if one neglects spin [36]. Thus, the Klein-Gordon equation was not a right way of bringing the principles of relativity and quantum theory together. Dirac's equation found the right way to do so by, mathematically, taking a square root of the Klein-Gordon equation, which may not be so difficult by current mathematical standards of theoretical physics, but was nontrivial at the time.[14]

---

[14] Einstein developed a major interest in Dirac's equation, as a spinor equation, and he used it, in his collaborations with W. Mayer, as part of his program for the unified field theory, conceived as a classical-like field theory, modeled on general relativity, and in opposition to quantum mechanics and, by then, quantum field theory. Accordingly, he only considered a classical-like spinor form of Dirac's equation, thus depriving it of (Einstein might have thought "freeing" it from) its quantum features, most fundamentally, discreteness ($h$ did not figure in Einstein's form of Dirac's equation), and probability. Einstein hoped but failed to derive discreteness from the underlying field-continuity. As noted above, by this point Einstein abandoned the principle approach in favor of the constructive approach, and his use on Dirac's equation was part of this new way of thinking. He was primarily interested in the mathematics of spinors, which he generalized in what he called "semivectors." While relevant, including in the context of the quantum-informational (principle) derivation of Dirac's equation in Section 5 [11], the subject is beyond my scope here. It is extensively discussed in [16]. It is worth noting that, unlike Einstein, O. Klein (for example, in his version of the Kaluza-Klein theory) always took quantum principles, especially discreteness, as primary, rather than aiming, as Einstein, to derive quantum discreteness from an underlying

Dirac's mathematical task was more difficult than that of Heisenberg because the conditions just outlined required both new variables, as in Heisenberg's scheme, and, in contrast to Heisenberg's scheme (which used the equations of classical mechanics), a new equation. In other words, as I said, Dirac didn't use the mathematical correspondence principle to derive his equation, but only to show that it converts into Schrödinger's equation at the quantum-mechanical limit. Dirac did, however, use the Klein-Gordon equation, of which he took a "square root," to satisfy the necessary principles of quantum mechanics and to give then a proper (relativistic) mathematical expression. As those of Heisenberg's matrix mechanics, Dirac's new variables proved to be noncommuting matrix-type variables, but of a more complex character, involving the so-called spinors and the multicomponent wave functions, the concept discovered by Pauli in his nonrelativistic theory of spin [36]. Just as Heisenberg's matrices earlier, Dirac's spinors had never been used in physics previously, although they were introduced in mathematics by W. C. Clifford about fifty years earlier (following the work of H. Grassmann on exterior algebras). And just as Heisenberg in the case of his matrices, Dirac was unaware of the existence of spinors and reinvented them.

In spite of the elegant simplicity of its famous compact form,

$$i\gamma \cdot \partial \psi = m\psi,$$

reproduced on the plate in Westminster Abbey commemorating Dirac, Dirac's equation encodes an extremely complex Hilbert-space machinery. The equation, as introduced by Dirac, was

$$(\beta mc^2 + \sum_{k=1}^{3} \alpha_k p_k c)\psi(x,t) = i\hbar \frac{\partial \psi(x,t)}{\partial t}$$

The new mathematical elements here are the 4×4 matrices $\alpha_k$ and $\beta$ and the four-component wave function $\psi$. The Dirac matrices are all Hermitian,

$$\alpha_i^2 = \beta^2 = I_4$$

($I_4$ is the identity matrix), and the mutually anticommute:

$$\alpha_i \beta + \beta \alpha_i = 0$$
$$\alpha_i \alpha_j + \alpha_j \alpha_i = 0$$

The above single symbolic equation unfolds into four coupled linear first-order partial differential equations for the four quantities that make up the wave function. The matrices form a basis of the corresponding Clifford algebra. One can think of Clifford algebras as quantizations of Grassmann's exterior algebras, in the same way that the Weyl algebra is a quantization of

---

continuity of a classical-like field theory. That is hardly surprising coming from a long-time assistant of Bohr. Klein's thinking, which led to several major contributions, was always quantum-oriented. It is just that the Klein-Gordon equation did not manage to bring quantum theory and relativity together successfully. The equation itself was later used in meson theory. Of course, Dirac's equation, too, was a unification of quantum mechanics and special relativity, albeit not of the kind Einstein wanted.

symmetric algebra. Here, *p* is the momentum operator in Schrödinger's sense, but in a more complicated Hilbert space than in standard quantum mechanics. The wave function $\psi$ *(t, x)* takes value in a Hilbert space $X = C^4$ (Dirac's spinors are elements of $X$). For each *t*, *y (t, x)* is an element of $H = L^2(R^3; X) = L^2(R^3) \otimes X = L^2(R^3) \otimes C^4$. This mathematical architecture allows one to predict the probabilities of quantum-electro-dynamical (high-energy) events, which, as explained below, have a greater complexity than quantum-mechanical (low-energy) events.

Finding new matrix-type variables or, more generally, Hilbert-space operators became the defining mathematical element of quantum theory. The current theories of weak forces, electroweak unifications, and strong forces (quantum chromodynamics) were all discovered by finding such variables. This is correlative to establishing the transformation group, a Lie group, of the theory and finding representations of this group in the corresponding Hilbert spaces. This is true for Heisenberg's matrix variables as well, as was discovered by H. Weyl and E. Wigner, for the Heisenberg group. In modern elementary-particle theory, irreducible representations of such groups correspond to elementary particles, the idea that was one of Wigner's major contributions to quantum physics [37]. This was, for example, how M. Gell-Mann discovered quarks, because at the time there were no particles corresponding to the irreducible representations (initially there were three of those, corresponding to three quarks) of the symmetry group of the theory, the so-called SO (3). It is the group of all rotations around the origin in the three-dimensional space, $R^3$, rotations represented by all three-by-three orthogonal matrices with determinant 1. (This group is noncommutative.) The electroweak group that Gell-Mann helped to find as well is SU (2), the group of two by two matrices with the determinant 1. Quarks are part of both theories. The genealogy of this group-theoretical thinking extends from Dirac's four by four matrices and, earlier, Pauli's two by two spin matrices. Gell-Mann famously borrowed the term quark from James Joyce's *Finnegans Wake*, which, as my epigraph suggests, might, in its notorious complexity, have been in turn inspired by quantum theory.

Dirac begins his paper by commenting on previous relativistic treatments of the electron, specifically the Klein-Gordon equation and its insufficiencies. He says:

> [The Gordon-Klein theory] appears to be satisfactory so far as emission and absorption of radiation are concerned, but is not so general as the interpretation of the non-relativi[stic] quantum mechanics, which has been developed sufficiently to enable one to answer the question: What is the probability of any dynamical variable at any specified time having a value laying between any specified limits, when the system is represented by a given wave function $\psi_n$? The Gordon-Klein interpretation can answer such questions if they refer to the position of the electron … but not if they refer to its momentum, or angular momentum, or any other dynamic variable. We would expect the interpretation of the relativi[stic] theory to be just as general as that the non-relativi[stic] theory. [1, pp. 611-612]

The term "interpretation" means here a mathematical representation of the physical situation, rather than, as is common now, a physical interpretation of a given quantum formalism cum the phenomena it relates to. Dirac's statement does not mean that a physical description of quantum processes in space and time is provided, as against only predictions, in general probabilistic, of the outcomes of quantum experiments. As is clear from this passage, Dirac thought the capacity of a given theory to enable such predictions sufficient if such predictions are possible for any dynamic variable. The main deficiency of the Klein-Gordon scheme was its inability to answer the following question, which indeed define all quantum theory as we understand it since quantum mechanics: "What is the probability of any dynamical variables at any specified time having a value laying between any specified limits, when the system is represented by a given wave function $\psi_n$?" Dirac then argues that the derivative first-order in time, missing in the Klein-Gordon equation, is a proper starting point for the relativistic theory of the electron. He says:

"The general interpretation of non-relativi[stic] quantum mechanics is based on the transformation theory, and is made possible by the wave equation being of the form

$$(H - W)\psi = 0, \qquad (1)$$

*i.e.*, being linear in W or $\frac{\partial}{\partial t}$, so that the wave function at any time determines the wave function at any later time. The wave function of the relativi[stic] theory must also be linear in W if the general interpretation is to be possible" [1, p. 612].

Before proceeding to his derivation, Dirac comments, in the statement with which I began here, on the second difficulty of the Klein-Gordon equation, that of the transitions from states of positive energy to those of negative energy. Dirac's theory inherits this difficulty because mathematically every solution of Dirac's equation is a solution of the Klein-Gordon equation, of which Dirac's equation is a square root. (The opposite is not true.) Luckily for the future of quantum theory, the difficulty proved to be not the weakness but the strength of Dirac's theory.

Dirac's derivation of his equation follows two key principles. The first is the invariance under Lorentz transformations (a relativity principle). The second is the mathematical correspondence principle: the equivalence of whatever the new equation one finds to Schrödinger's equation (equation (1) above) in the limit of large quantum numbers, which requires correspondence with Pauli's spin theory as an intermediate step [1, p. 613]. Other key quantum principles are fulfilled and given their mathematical expression automatically once this correspondence, again, lacking in the Klein-Gordon theory, is in place.

In the absence of the external field, which Dirac considers first and to which I shall restrict myself here, since it is sufficient for my main argument, the Klein-Gordon equation "reduces to

$$(-p_0^2 + \mathbf{p}^2 + m^2 c^2)\psi = 0 \quad (3)$$

if one puts

$$p_0 = \frac{W}{c} = i\frac{\hbar}{c}\frac{\partial}{\partial t}$$"

[1, p. 613]

Next Dirac uses the symmetry between time, $p_0$, and space, $p_1, p_2, p_3$, required by relativity, which implies that because the Hamiltonian one needs is linear in $p_0$, "it must also be linear in $p_1$, $p_2$, and $p_3$." He then says:

[the necessary] wave equation is therefore in the form

$$(p_0 + \alpha_1 p_1 + \alpha_2 p_2 + \alpha_3 p_3 + \beta)\psi = 0 \quad (4)$$

where for the present all that is known about the dynamical variables or operators $\alpha_1$, $\alpha_2$, $\alpha_3$, and $\beta$ is that they are independent of $p_0, p_1, p_2, p_3$, i.e., that they commute with $t, x_1$, $x_2, x_3$. Since we are considering the case of a particle moving in empty space, so that all points in space are equivalent, we should expect the Hamiltonian not to involve $t, x_1, x_2$, $x_3$. This means that $\alpha_1, \alpha_2, \alpha_3$, and $\beta$ are independent of $t, x_1, x_2, x_3$, i.e., that they commute with $p_0, p_1, p_2, p_3$. We are therefore obliged to have other dynamical variables besides the co-ordinates and momenta of the electron, in order that $\alpha_1, \alpha_2, \alpha_3, \beta$ may be functions of them. The wave function $\psi$ must then involve more variables than merely $x_1$, $x_2, x_3, t$.

Equation (4) leads to

$$0 = (-p_0 + \alpha_1 p_1 + \alpha_2 p_2 + \alpha_3 p_3 + \beta)(p_0 + \alpha_1 p_1 + \alpha_2 p_2 + \alpha_3 p_3 + \beta)\psi =$$
$$[-p_0^2 + \Sigma\alpha_1^2 p_1^2 + \Sigma(\alpha_1\alpha_2 + \alpha_2\alpha_1)p_1 p_2 + \beta^2 + \Sigma(\alpha_1\beta + \beta\alpha_1)]\psi$$

(5)

where $\Sigma$ refers to cyclic permutation of the suffixes 1, 2, 3. [1, p. 613]

I pause here to reflect on Dirac's way of thinking, manifested in this passage and throughout his derivation of his equation. Taking advantage of noncommutativity in (5) is worth a special notice not only because this is one of Dirac's forte but also because, as Dirac was, as I said, among the first to realize, it represents the mathematical essence of quantum theory and is crucial to the mathematical expression of its fundamental principles. Equation (4), a square root of (3), the Klein-Gordon equation, is already Dirac's equation in abstract algebraic terms, thus, expressing Dirac's approach of finding, first, an abstract mathematical scheme suitable for the expression of the principles of quantum theory and, correlatively, for predicting the data in question in his relativistic theory of the electron. One will now need to find $\alpha_n$ and $\beta$, to find the actual form of the equation. Inspired by that of Heisenberg as it was, Dirac's approach is different from that of Heisenberg. While mathematics and the invention of a cohesive formal mathematical scheme are no less important for Heisenberg, one does not find in Heisenberg the same kind of, to use Dirac's word, *play* with (still more) abstract structures that is characteristic of Dirac [35]. Dirac's earlier work on *q*-numbers quantum-mechanical formalism already displayed this power of abstract mathematical thinking, again, however, arising from and governed by physical principles of quantum theory. It is true that antimatter was a consequence of the mathematical structure of Dirac's equation. This is not uncommon in theoretical physics, whether constructive or principle: new physical objects are discovered and new physical principles are often established as consequences of the mathematical formalism. This is what happened in the case of Dirac's equation as well. Nevertheless, Dirac's equation was what it was because of the fundamental physical principles on which it was based. It is equally crucial, however, that this equation mathematically expressed these principles. Dirac now proceeds as follows:

[Equation (5)] agrees with (3) [the Klein-Gordon equation, in the absence of the external field $(-p_0^2 + \mathbf{p}^2 + m^2 c^2)\psi$] if

$$\alpha_r = 1, \alpha_r\alpha_s + \alpha_s\alpha_r = 0 \ (r \neq s) \ r, s = 1,2,3.$$
$$\beta^2 = m^2 c^2, \qquad \alpha_r\beta + \beta\alpha_r = 0$$

If we put $\beta = \alpha_4 mc$, these conditions become

$$\alpha_\mu^2 = 1, \alpha_\mu\alpha_\nu + \alpha_\nu\alpha_\mu = 0 \ (\mu \neq \nu)$$
$$\mu, \nu = 1, 2, 3, 4. \qquad (6)$$

[1, p. 613]

Dirac, again, takes advantage of a partial mathematical correspondence with the Klein-Gordon equation (that between the function of a complex variable and its square root), which allows him to derive certain algebraic conditions that $\alpha_\mu$ and $\beta$ must satisfy. Dirac will now state that "we can suppose $\alpha_\mu$'s to be expressed in some matrix scheme, the matrix elements of $\alpha_\mu$ being, say,$\alpha_\mu$ $(\zeta'\zeta'')$" [1, p. 613]. This supposition is not surprising given both the formal mathematical considerations (such as the anticommuting relations between them) and the preceding history of matrix mechanics, including Dirac's own previous work. We know or may safely assume from Dirac's account of his work on his equations that matrix manipulation, "playing with equations," as he called it, was one of his starting points [35]. In addition, Pauli's theory, which is about to enter Dirac's argument, provided a ready example of a matrix scheme [36; 38, pp. 55-56, 60]. It was clear that matrix algebra of some sort is a good candidate for $\alpha_\mu$. Dirac, again, gets extraordinary mileage from considering the formal properties of the variables involved, even before considering what these variables actually are and as a way of gauging what they should be, which is his next step, which reveals another remarkable consequence of the necessity of the particular matrix variables required by Dirac. For if we "suppose $\alpha_\mu$'s to be expressed in some matrix scheme, the matrix elements of $\alpha_\mu$ being, say, $\alpha_\mu$ $(\zeta'\zeta'')$," then "the wave function $\psi$ must be a function of $\zeta$ as well as $x_1, x_2, x_3, t$. The result of $\alpha_\mu$ multiplied into $\psi$ will be a function $(\alpha_\mu, \psi)$ of $x_1, x_2, x_3, t, \zeta$ defined by

$$(\alpha_\mu, \psi)(x, t, \zeta) = \Sigma_{\zeta'}\, \alpha_\mu\,(\zeta\zeta')\psi\,(x, t, \zeta')" \ [1, \text{p. 614}].$$

Dirac is now prepared "for finding four matrices $\alpha_\mu$ to satisfy the conditions (6)," those forming the Clifford algebra, and for finding the actual form of variables that satisfy formal equation (4) or (5). Dirac considers first the three Pauli spin matrices, which satisfy the conditions (6), but not the equations (4) or (5), which needs four by four matrices. He says:

We make use of the matrices

$$\sigma_1 = \begin{pmatrix} 0 & 1 \\ 1 & 0 \end{pmatrix};\ \sigma_2 = \begin{pmatrix} 0 & -i \\ i & 0 \end{pmatrix};\ \sigma_3 = \begin{pmatrix} 1 & 0 \\ 0 & 1 \end{pmatrix}$$

which Pauli introduced to describe the three components of the spin angular momentum. These matrices have just the properties

$$\sigma_r^2 = 1\ \ \sigma_r\sigma_s + \sigma_s\sigma_r = 0\ (r \neq s), \qquad (7)$$

that we require for our $\alpha$'s. We cannot, however, just take the $\sigma$'s to be thereof our $\alpha$'s, because then it would not be possible to find the fourth. We must extend the s's in a diagonal matter to bring in two more rows and columns, so that we can introduced three more matrices $r_1, r_2, r_3$ of the same form as $s_1, s_2, s_3$, but referring to different rows and columns, thus:

$$\rho_1 = \begin{pmatrix} 0 & 0 & 1 & 0 \\ 0 & 0 & 0 & 1 \\ 1 & 0 & 0 & 0 \\ 0 & 1 & 0 & 0 \end{pmatrix};\ \rho_2 = \begin{pmatrix} 0 & 0 & -i & 0 \\ 0 & 0 & 0 & -i \\ i & 0 & 0 & 0 \\ 0 & i & 0 & 0 \end{pmatrix};\ \rho_3 = \begin{pmatrix} 1 & 0 & 0 & 0 \\ 0 & 1 & 0 & 0 \\ 0 & 0 & -1 & 0 \\ 0 & 0 & 0 & -1 \end{pmatrix};$$

$$\sigma_1 = \begin{pmatrix} 0 & 1 & 0 & 0 \\ 1 & 0 & 0 & 0 \\ 0 & 0 & 0 & 1 \\ 0 & 0 & 1 & 0 \end{pmatrix}; \ \sigma_2 = \begin{pmatrix} 0 & -i & 0 & 0 \\ i & 0 & 0 & 0 \\ 0 & 0 & 0 & -i \\ 0 & 0 & i & 0 \end{pmatrix}; \ \sigma_3 = \begin{pmatrix} 1 & 0 & 0 & 0 \\ 0 & -1 & 0 & 0 \\ 0 & 0 & 1 & 0 \\ 0 & 0 & 0 & -1 \end{pmatrix};$$

The ρ's are obtained from σ's by interchanging the second and the third row, and the second and the third columns. We now have, in addition to equations (7)

$$\rho_r^2 = 1 \quad \rho_r \rho_s + \rho_s \rho_r = 0 \ (r \neq s),$$

and also

$$\rho_r \sigma_t = \sigma_t \rho_r \ (7')$$

[1, p. 615]

These matrices are Dirac's great invention, parallel to Heisenberg's invention of his matrix variables. Dirac's matrices form the basis of the corresponding Clifford algebra and define the mathematical architecture where the multicomponent relativistic wave function for the electron appears from the first principles, in contrast to Pauli's theory where the two-component nonrelativistic wave function, necessary to incorporate spin, appears phenomenologically. Spin is an automatic consequence of Dirac's theory, unintended (just as antimatter was), but fundamental. The entities Dirac's matrices transform are different from either vectors or tensors and are called spinors, introduced, as mathematical objects, by E. Cartan (who did not use the term itself initially) in 1913.

The rest of the derivation of Dirac's equation is a nearly routine exercise, with a few elegant but easy matrix manipulations. Dirac also needs to prove the relativistic invariance and the conservation of the probability current, and to consider the case of the external field, none of which is automatic, but is standard textbook material at this point. The most fundamental and profound aspects of Dirac's thinking as principle thinking are contained in the parts of his paper just discussed, and I can draw my main conclusions concerning the significance and implications of Dirac's *theory* of the relativistic electron, which his equation embodies, from them.

Dirac's theory is a remarkable example of principle thinking in theoretical physics, in which fundamental physical principles of relativity and quantum theory combine with mathematics so that physics guides the mathematics, which both gives these principles their precise mathematical expression and leads to new physics. Dirac's theory was not only a result of his confidence in already established principles, but also led to new principles, inherent or implied in his equation, although it took a while to realize the radical nature of these implications. The most crucial of them emerged from the concepts of antimatter, the most revolutionary consequence of Dirac's theory (although the experimental discovery of the positron by C. D. Anderson, in 1932, was independent), in particular what may be called the particle-transformation principle, the PT principle. This principle extends and is a consequence of the antiparticle principle, which states that for every particle there is an antiparticle, although some particles, sometimes known as Majorana particles, such as photons, are their own antiparticles. The PT principle adds the loss of particle identity in quantum experiments to the nonrealism of the RWR principle found already in quantum mechanics, if interpreted in the spirit of Copenhagen. I shall now discuss the PT principle, which changed our concept of the elementary particle. It is also correlative to various symmetry and invariance principles of quantum theory.

As explained above, Dirac's equation encodes a complex mathematical architecture, which manifests the fundamental physical principles of quantum electrodynamics and by implication quantum field theory. (Dirac's equation is not quite a field equation, but given its essentially quantum-field-theoretical physical nature, it would also be difficult to see it in terms of relativistic quantum mechanics, as some suggest (e.g., [39]). The Hilbert space associated with a given quantum system in Dirac's theory is a tensor product of the infinite dimensional Hilbert space (encoding the mathematics of continuous variables) and a finite-dimensional Hilbert space over complex numbers, which, in contrast to the two-dimensional Hilbert space of Pauli's theory, $C^2$, is four-dimensional in Dirac's theory, $C^4$. (Spin is contained in the theory automatically.) Dirac's wave function $\psi\,(t,\,\boldsymbol{x})$ takes value in a Hilbert space $X = C^4$ (Dirac's spinors are elements of $X$). For each $t$, $\psi\,(t,\,\boldsymbol{x})$ is an element of

$$H = L^2\,(R^3;\,X) = L^2\,(R^3) \otimes X = L^2\,(R^3) \otimes C^4.$$

Other forms of quantum field theory give this type of architecture an even greater complexity. It is difficult to overestimate the significance of this architecture, which amounts to a very radical view of matter, first manifested in the existence of antimatter. This architecture mathematically responds to, and led to a discovery of, the following physical situation, keeping in mind that, as quantum mechanics, Dirac's equation or, more generally, quantum field theory only provides probabilities for the outcomes of quantum events, registered in measuring instruments.

Suppose that one arranges for an emission of an electron, at a given high energy, from a source and then performs a measurement at a certain distance from that source. Placing a photographic plate at this point would do. The probability of the outcome would be properly predicted by quantum electrodynamics. But what will be the outcome? The answer, as we know, is not what a classical or even our quantum-mechanical intuition would expect, and this unexpected answer was a revolutionary discovery of quantum electrodynamics, beginning with Dirac's equation. To appreciate the revolutionary nature of this discovery, let us consider, first, what happens if we deal with a classical object, analogous to an electron, and then a low-energy quantum electron in the same type of arrangement. I speak of a classical object because the "game of small marbles" for electrons was finished even before quantum mechanics, because an electron would be torn apart by the force of its negative electricity. This required theoretical physics to treat electron mathematically as a dimensionless point, without really giving it a physical concept, as Dirac does not fail to note in his paper, in conjunction with spin, which obviously complicated the situation [1, p. 610]. One could still treat an electron classically, for example, by the correspondence principle, when an electron is far away from the nucleus. This treatment is an idealization because this behavior is quantum, and hence could lead to quantum effects described below. On the other hand, within the idealization of classical physics, we may treat classical objects (Newton did so already) as dimensionless point endowed with mass.

We can take as a model of the classical situation a small ball that hits a metal plate, which can be considered as either a position or a momentum measurement, or indeed a simultaneous measurement of both, and time $t$. In classical mechanics we can deal directly with the objects involved, rather than with their effects upon measuring instruments. The place of the collision could, at least in an idealized representation of the situation, be predicted exactly by classical mechanics, and we can repeat the experiment with the same outcome on an identical or even the same object. Most importantly, regardless of where we place the plate, we always find the same object, at least in an experimental situation shielded from outside interferences, which could deflect the ball or even destroy it before it reaches the plate.

By contrast, if we deal with an electron in the quantum-mechanical regime, beyond the fact that it is impossible, because of the uncertainty relations, to predict the place of collision exactly or with the degree (in principle unlimited) of approximation possible in classical physics, there is a nonzero probability that we will not observe such a collision at all. It is also not possible to distinguish two observed traces as belonging to two different objects of the same type, or to distinguish such objects in the first place, a circumstance that becomes even more crucial in high-energy regimes. Unlike in the classical case, in dealing with quantum objects, there is no way to improve the conditions of the experiment to avoid this situation. Quantum mechanics, however, gives us correct probabilities for such events. This is accomplished by defining the corresponding Hilbert space, with the position and other operators as observables, and using the formalism, say, Schrödinger's equation and Born's rule for obtaining the statistics of possible outcomes, once we repeat the experiment a large number of times. In a single experiment an electron could, in principle, be found anywhere, or, again, not found at all.

Once the process occurs at a high energy and is governed by quantum electrodynamics, the situation is still different, indeed radically different. One might find, in the corresponding region, not only an electron, as in classical physics, or an electron or nothing, as in the quantum-mechanical regime, but also other particles: a positron, a photon, an electron–positron pair. Just as does quantum mechanics, quantum-electrodynamics, beginning with Dirac's equation, rigorously predicts which among such events can occur, and with what probability, and, in the present view it can only predict such probabilities, or statistical correlations between certain quantum events. In order to do so, however, the corresponding Hilbert-space machinery becomes much more complex, essentially making the wave function $\psi$ a four-component Hilbert-space vector, as opposed to a one-component Hilbert-space vector, as in quantum mechanics. This Hilbert space is, as noted, $H = L^2(R^3; X) = L^2(R^3) \otimes X = L^2(R^3) \otimes C^4$ and the operators are defined accordingly. This structure allows for a more complex structure of predictions (which are still probabilistic) corresponding to the situation just explained, usually considered in terms of virtual particle formation and Feynman's diagrams. Once we move to still higher energies or different domains governed by quantum field theory the panoply of possible outcomes becomes much greater. The Hilbert spaces and operator algebras involved would be given a yet more complex structure, in relation to the appropriate Lie groups and their representations, defining (when these representations are irreducible) different elementary particles, as indicated above [37]. In the case of Dirac's equation we only have electron, positron, and photon.

It follows that in quantum field theory an investigation of a particular type of quantum object irreducibly involves not only other particles of *the same type* but also *other types* of particles. The identity of particles within each type is strictly maintained in quantum field theory, as it is in quantum mechanics. In either theory one cannot distinguish different particles of the same type, such as electrons. According to H. Weyl, "the possibility that one of the identical twins Mike and Ike is in the quantum state E1 and the other in the quantum state E2 does not include two differentiable cases which are permuted on permuting Mike and Ike; it is impossible for either of these individuals to retain his identity so that one of them will always be able to say 'I'm Mike' and the other 'I'm Ike.' Even in principle one cannot demand an alibi of an electron!"[40, p.

241].[15] One cannot be certain that one encounters the same electron in the experiment just described even in the quantum-mechanical situation, although the probability that it would be a different electron is low. In quantum field theory, it is as if instead of identifiable moving objects and motions of the type studied in classical physics, we encounter a continuous emergence and disappearance, creation and annihilation, of particles, theoretically governed by the concept of virtual particle formation. (This description is still too crude and one needs to supplement it by the concept of quantum field, but it will suffice for making my main point here.) The operators used to predict the probability of such events are the creation and annihilation operators.

This view clearly takes us beyond quantum mechanics. For, while the latter questions the applicability of classical concepts, such as objects (particles or waves) and motion, at the quantum level, it still preserves the identity of quantum objects and of the types of quantum objects within the same experiment. It is still possible to speak of this identity, even though, in the present view, these objects themselves remain unthinkable and only manifest themselves and their identity to each other (the type of their identity, say, electrons vs. photons) in their effects upon measuring instruments. This is no longer possible, even within the same experiment, in the quantum-field-theoretical regimes, because, as just explained, one may, in the course of the same experiment observe different types of particles, which leads to the particle transformation (PT) principle. It is a principle because this situation is found and, due to this principle qua principle, is to be expected in any regime governed by quantum field theory. This principle was at work, in conjunction with or as correlative to various symmetry principles, in the quantum field theory of nuclear forces, for example, and governed the practice of theoretical physics, not the least in many discoveries of new particles, such as quarks. Indeed, in conjunction with or as correlative to various symmetry and invariance principles, but as a *more expressly* physical one, this principle is one of the most crucial principles of high-energy theoretical physics.[16]

The emergence of this situation and this principle and this set of principles with Dirac's theory or in the wake of it and as its development was a momentous event in the history of quantum physics, comparable to that of Heisenberg's introduction of his matrix variables, or even more momentous, according to Heisenberg. He saw Dirac's theory as an even more radical revolution than quantum mechanics was. In his article, revealingly entitled "What is an Elementary Particle?" and devoted to the role of symmetry principles in particle physics, Heisenberg spoke of Dirac's discovery of antimatter as "perhaps the biggest change of all the big changes in physics of our century … because it changed our whole picture of matter. … It was one of the most spectacular consequences of Dirac's discovery that the old concept of the elementary particle collapsed completely" [43, pp. 31–33].

A path to a new understanding of the ultimate constitution of nature became open, however. Although still unanswered, the question of Heisenberg's title was advanced, as question, immeasurably by Dirac's equation and then quantum electrodynamics and quantum field theory, and was advanced still much further since the time of Heisenberg's article in the 1970s, reaching the so-called Standard Model of particle physics, which is quantum-field-theoretical.[17] This advancement was essentially guided by the same principles. Following this path, quantum field theory made remarkable progress since its introduction or, again, since Heisenberg's remark, a progress resulting in the electroweak unification and the quark model of nuclear forces, developments that commenced around the time of these remarks. Many predictions of the theory, from quarks to electroweak bosons and the concept of confinement and asymptotic freedom, to name just a few, were spectacular, and the field has garnered arguably the greatest number of Nobel Prizes in physics. It was also quantum field theory that led to string and then brane theories, the current stratosphere and the site of new controversies of theoretical physics.

---

[15] The statement is cited in [41], which considers the question of identity and indistinguishability of elementary particles (i.e., their indistinguishability from each other within the same particle type, e.g., electrons vs. photons) from a realist perspective. See also [42], for a comprehensive realist treatment of the subject.

[16] Although some of these principles are mathematical, they reflect and often express profound physical principles, as does, for example, the gauge symmetry principle, found already in Maxwell's electrodynamics, but especially important in general relativity and quantum field theory, as well as in most proposals for quantum gravity. Thus, quantum electrodynamics is an abelian gauge theory with the symmetry group U(1) (this group is commutative), and it has one gauge field, with the photon being the gauge boson. The Standard Model is a non-abelian gauge theory with the symmetry group U(1)×SU(2)×SU(3) and broken symmetries, and it has a total of twelve gauge bosons: the photon, three weak bosons, and eight gluons.

This is not to say that the theory is free of difficulties, even apart from it mathematically unwieldy and as yet incomplete character (vis-à-vis quantum mechanics within its proper limit). The standard model is only partially unified thus far (there is no single symmetry group), although achieving such a complete unification is not always seen, including by this author, as imperative, insofar as the theories involved predict all the available data in their respective domains. The theory's unwieldiness, although much bemoaned by Dirac, is not necessarily a problem either, insofar as it works. What are, then, the problems? Those that led to the necessity of renormalization might be. These problems had begun to emerge from the early 1930s, when it was realize that the computations provided by quantum electrodynamics were reliable only as a first order of perturbation theory and led to the appearance of infinities or divergences in the theory once one attempted to make calculations that would provide closer approximations matching the experimental data. These difficulties were eventually handled through the renormalization procedure, which became and has been ever since a crucial part of the machinery of quantum electrodynamics and quantum field theory. In the case of quantum electrodynamics, renormalization was performed in the later 1940s by S-I. Tomonaga, J. Schwinger, and R. Feynman (which brought them a joint Nobel Prize in 1965), with important contributions by others, especially F. Dyson, and earlier H. Bethe and H. Kramers. The Yang-Mills theory, which grounds the standard model, was eventually shown to be renormalizable by M. Veltman and G. 't Hooft in the 1970s (bringing them their Nobel Prize). This allowed a proper development of the Standard Model of all forces of nature, except for gravity, which, unlike others, has not been given its quantum form thus far.

The renormalization procedure is difficult mathematically even in quantum electrodynamics (the mathematics of the electro-weak theory or of quantum chromodynamics, which handles the strong force, is nearly prohibitively difficult). While it is possible to see renormalization in more benign ways (e.g., via the so-called "renormalization group" and "effective quantum field theories") and while it has been very effective thus far, its mathematical legitimacy is still under cloud. Roughly speaking, the procedure might be seen as manipulating infinite integrals that are divergent and, hence, mathematically illegitimate. At a certain stage of calculation, however, these integrals are replaced by finite integrals through artificial cut-offs that have no proper mathematical justification within the formalism and are performed by putting in, by hand, experimentally obtained numbers that make these integrals finite, which removes the infinities from the final results of calculations. These calculations are experimentally confirmed to a very high degree. I will not address the details of renormalization further, and will allow myself to refer to [45; 46; 47, pp 149-168] and, for a historical account, to [47, pp. 595-605]. The discussion of the relevant subsequent developments, such as the renormalization group, effective quantum field theories, and so forth is also beyond my scope here, but are considered in [39, 40].

---

[17] The title was reprised by S. Weinberg's 1996 article, reflecting on a more advance stage of quantum field theory, without, however, answering the question either [44].

Renormalization or other difficulties just mentioned need not mean that quantum field theory should be replaced by a different theory, although Dirac thought so throughout his life and he wasn't alone. Luckily, renormalization has worked thus far. Quantum electrodynamics is the best experimentally confirmed theory thus far, and many predictions of quantum field theory beyond quantum electrodynamics have been spectacular. The discovery of Higgs boson, an essential component of the Standard Model, is the latest example, although some of the earlier discoveries, such as those of the electroweak (W+, W−, and Z) bosons, the top quark, and the tau-lepton, are hardly less significant. The Higgs discovery may still require further confirmation, and the data involved may require advancing and perhaps adjusting the Standard Model itself, which is, again, not entirely complete in any event [48]. However, the discovery was deemed confirmed enough (two independent detectors indicated a likely presence of the Higgs particle) to award the 2013 Nobel Prize to F. Englert and P. Higgs, who were among those involved in the development of the mathematical theory of the Higgs field. Our future fundamental theories might prove to be finite (some versions of string and brane theory appear to hold such a promise), thus proving that the necessity of renormalization is merely the result of the limited reach of our quantum theories at present. (Effective quantum field theories are based on this view.) The emergence of other finite alternatives, proceeding along entirely different trajectories, may not be excluded either. While, however, a finite theory may be preferable, renormalization may not be a very big price to pay for the theory's extraordinary capacity to predict the increasingly complex manifold of quantum phenomena that physics has confronted throughout the history of quantum field theory. And then, that a finite theory will not be found cannot be excluded either, in which case renormalization will continue to be our main hope.

The theory does remain incomplete insofar it does not cover the scales beyond the standard scale of high-energy physics, where in particular the effects of gravity would need to be taken into account. String or brane theory is still the most widely entertained proposal, born from and closely and fundamentally related to quantum field theory. The theory is even more complex mathematically than quantum field theory and remains highly hypothetical physically, with uncertain chances to be connected to experiments in any near future, although certain consequences of the theory could eventually be tested. These factors have made the theory controversial. While it has many prominent advocates and promoters, and a large (although no longer quite as large) cohort of practitioners and adherents, it also has quite a few detractors, sometimes to the point of dismissing it as useless and even "dangerous" metaphysics. But then, most proposed alternatives, such as, say, loop quantum gravity, are hardly less hypothetical or controversial. I shall not enter into these controversies here. Instead, I would like, in closing this article to consider a possible alternative approach to high-energy quantum physics, an approach, based in quantum information theory, by G. M. D'Ariano and coworkers [9, 10, 11].

## 5. Principles of Information Processing, Quantum Theory, and Dirac's Equation

While string or brane theory or most other current proposals for physics beyond quantum field theory are constructive theories, D'Ariano and coworkers' approach to Dirac's equation is a principle one and, as such, is closer to the spirit of Heisenberg's and Dirac's work, as considered here. It has a constructive dimension as well, introduced by the idea of quantum cellular automata (QCA), but this aspect of their framework will not be considered here [9]. In addition, the fact that Dirac's equation can be derived strictly from "principles of information processing," as embodied in the QCA, without using relativity, suggests, at least to this writer, that the ultimate architecture of matter may be quantum. If so, relativity, at least special relativity (but, not inconceivably, general relativity as well), is a surface-scale effect of this quantumness, although the overall set of principles of information processing in question extends beyond those defining this quantumness. Admittedly, the approach is at an early stage, and thus far it only offers a derivation of Dirac's equation, which, however, is a major step in extending quantum information theory beyond quantum mechanics. I cannot, within my scope, consider the technical aspects of this derivation and also have to bypass several key features of the authors' physical argument, including the role of quantum automata in their framework, developed in several articles, to which and further references there I must refer the reader for further details [9, 10, 11]. I shall only comment on the role of informational principles in these works. I shall consider first the authors' program of finite-dimensional quantum *theory* based in these principles, and then comment on their derivation of Dirac's equation.[18]

The program is inspired in part by "a need for a *deeper understanding of quantum theory* itself from fundamental principles" (which, the authors contend, has never been really achieved) and is motivated by the development of quantum information theory, and in part for that reason deals with discrete variables and the corresponding (finite-dimensional) Hilbert spaces.[19] According to the authors: "[T]he rise of quantum information science moved the emphasis from logics to information processing. The new field clearly showed that the mathematical principles of quantum theory imply an enormous amount of information-theoretic consequences. … The natural question is whether the implication can be reversed: is it possible to retrieve quantum theory from a set of purely informational principles?" [9, p. 1]. In contrast to several preceding attempts along quantum-informational lines during the last decade of so, the authors aim to offer "a complete derivation of finite-dimensional quantum theory based on purely operational informational principles:"

> In this paper we provide a complete derivation of finite dimensional quantum theory based on purely operational principles. Our principles do not refer to abstract properties of the mathematical structures that we use to represent states, transformations, or measurements, but only to the way in which states, transformations, and measurements combine with each other. More specifically, our principles are of *informational* nature: they assert basic properties of information processing, such as the possibility or impossibility to carry out certain tasks by manipulating physical systems. In this approach the rules by which information can be processed determine the physical theory, in accordance with Wheeler's program "it from bit," for which he argued that "all things physical are information-theoretic in origin" [51]. Note [however, that] our axiomatization of quantum theory is relevant, as a rigorous result, also for those who do not share Wheeler's ideas on the

---

[18] "Theory" here refers primarily to the mathematical structure of quantum theory, rather than to its mechanical or dynamical aspects, such as, in the case of the finite-dimensional quantum theory, found the quantum mechanics of discrete variables (spin). See Note 1 above.

[19] Among the key predecessors here are C. Fuchs's work, which, however, more recently "mutated" to a somewhat different program, that of quantum Bayesianism or QBism (e.g., [49]), and L. Hardy [50], equally motivated by the aim of deriving quantum mechanics from a more natural set of principles, postulates, or axioms. Hardy's paper was, arguably, the first rigorous derivation of that type. Neither of these two approaches is constructive or realist, nor, again, is that of D'Ariano et al. See [9] for further references. The different terms just mentioned (all of them are use by D'Ariano et al as well) do not affect the essential aspects of the programs in question at the moment. All these attempts refer to finite-dimensional quantum theories. Let me add that, while emphasizing the role of fundamental principles in quantum theory, the present article does not claim that a sufficient understanding of quantum theory itself, say, quantum mechanics, from such principles has been achieved. This remains an open question, even more so when dealing with continuous variables (to which my discussion has been restricted thus far), where the application of the principles of quantum information is more complex as well.

informational origin of physics. In particular, in the process of deriving quantum theory we provide alternative proofs for many key features of the Hilbert space formalism, such as the spectral decomposition of self-adjoint operators or the existence of projections. The interesting feature of these proofs is that they are obtained by manipulation of the principles, without assuming Hilbert spaces from the start. [9, p. 1][20]

One of the principles advanced by the authors, the purification principle, plays a particularly, indeed uniquely, important role in their program, as an essentially quantum principle:

The main message of our work is simple: within a standard class of theories of information processing, quantum theory is uniquely identified by a single postulate: *purification*. The concrete realization of Schrödinger's introduced in Ref. [10], expresses a distinctive feature of quantum theory, namely that the ignorance about a part is always compatible with the maximal knowledge of the whole. The key role of this feature was noticed already in 1935 by Schrödinger in his discussion about entanglement [30], of which he famously wrote "I would not call that *one* but rather *the* characteristic trait of quantum mechanics, the one that enforces its entire departure from classical lines of thought." In a sense, our work can be viewed as the concrete realization of Schrödinger's claim: the fact that every physical state can be viewed as the marginal of some pure state of a compound system is indeed the key to single out quantum theory within a standard set of possible theories. It is worth stressing, however, that the purification principle assumed in this paper includes a requirement that was not explicitly mentioned in Schrödinger's discussion: if two pure states of a composite system AB have the same marginal on system A, then they are connected by some reversible transformation on system B. In other words, we assume that all purifications of a given mixed state are equivalent under local reversible operations. [9, p. 2]

The authors also speak of "the purification postulate," and they refer to the remaining informational principles as "axioms," because "as opposed to the purification 'postulate,' … they are not at all specific [to] quantum theory" [9, p. 3]. While these terminological distinctions, especially the second one, may be somewhat tenuous, they do not affect the authors' argument. These postulates and axioms do define principles in the present sense (and principles often involve postulates), and in particular, jointly, provide the *guidance* for deriving the finite-dimensional quantum theory. Besides, as will be seen presently, the authors state their strictly operational principles later in the article [9, p. 6].

The purification principle is a new principle, although it has its genealogy in the previous operational approaches mentioned above, which, in particular, equally stress the (axiomatic) significance of the tensor product structure. First of all, even beyond the fact that it has a richer content than that of Schrödinger's statement, Schrödinger never saw his claim as a principle, perhaps also because of his critical view of quantum mechanics, in agreement with Einstein. The principle could be related, along the lines discussed earlier, to Bohr's RWR principle and his concept of completeness, as combined with complementarity, which implies that "the ignorance about a part [one of the two complementary parts] is always compatible with the maximal knowledge of the whole." Bohr saw the EPR experiment (the background for Schrödinger's claim and for his concept of entanglement, the term he introduced as well in German [*Verschränkung*] and English) as a manifestation of complementarity, as well as the RWR principle [27; 24, v. 2, p. 59].[21] The purification principle is, however, given a more rigorous mathematical expression (at least in the finite-dimensional case) than one finds in Bohr, whose work, as noted above, was not especially concerned with finding mathematical expressions for his key principles.

While having an essential and even unique role in the authors' operational derivation of the finite-dimensional quantum theory, the purification principle is not sufficient to do so. The authors need five additional axioms, which I shall state below. This is not surprising. As discussed above, in Heisenberg's main grounding quantum physical principles—the suspension of the description of quantum objects and processes (the proto-RWR principle) and the quantum-probability (QP) principle—were not sufficient to derive quantum mechanics. To do so, he needed the correspondence principle, to which he gave a mathematical form. Dirac, too, needed, a larger set of principles, postulates, and assumptions (even apart from those of relativity) for deriving his equation than those that define its specifically quantum character, including, again, a form of mathematical correspondence principle. What is remarkable, however, that one needs only one "postulate" to distinguish classical and quantum information theory. A similar situation transpires in Hardy's paper mentioned above, where indeed this difference turn not only on a single "axiom," but on the use of a single word, "continuity," technically a single feature of the situation, "the *continuity* of a reversible transformation between any two pure states" [50, p. 2; emphasis added].[22] It appears, in fact, that this situation reflects the roles of both complex numbers and the tensor product, in other words, that of complex Hilbert spaces or their mathematical equivalents in quantum theory, roles that appear uncircumventable.

---

[20] References inside this and other passages cited from [9], [10] and [11] are adjusted to follow the numbering of references in the present article.

[21] J. Bub's article, cited earlier, also considers quantum mechanics as a principle theory in order to account for the EPR-type experiments and quantum entanglement [6].

There are instructive specific parallels (not identical features!) between the authors' and Heisenberg's approaches, in particular between Heisenberg's proto-RWR principle and the purification principle. The QP principle present in both cases, given that D'Ariano et al (rightly) see quantum mechanics an "operational-probabilistic theory" of a special type, defined by the purification postulate [9, p. 3]. As they say: "The operational-probabilistic framework combines the operational language of circuits with the toolbox of probability theory: on the one hand experiments are described by circuits resulting from the connection of physical devices, on the other hand each device in the circuit can have classical outcomes and the theory provides the probability distribution of outcomes when the devices are connected to form closed circuits (that is, circuits that start with a preparation and end with a measurement)" [9, p. 3]. This is close to Heisenberg's and Bohr's view of the quantum-mechanical situation, keeping in mind the difference defined by the concept of "circuit" (not found in either Heisenberg or Bohr), on which I shall comment presently. As explained in Section 3, Heisenberg found his formalism by using the mathematical correspondence principle, not exactly the first principle, because it depended on the equations of classical mechanics at the classical limit where $h$ could be neglected. However, Heisenberg needed new variable because the classical variables (as functions of real variables) do not give Bohr's frequencies rules for spectra. Heisenberg discovered that these rules are satisfied by, in general, noncommuting matrix variables with complex coefficients, related to amplitudes, from which one derives, in essence by means of Born's rule for this case, the probabilities (or probability distributions) for transitions between stationary states (no longer assumed to be orbits) defining spectra, which are observed in measuring devices.

By contrast, D'Ariano et al arrive at the tensor-product architecture in a more first-principle-like way, in particular, independently of classical physics. (The latter, to begin with, does not have discrete variables, such as spin, which are purely quantum, with which the finite-dimensional quantum theory could be associated.) This is accomplished by using the rules governing the structure of operational devices, rules that are more empirical, albeit not completely, because they are given a mathematical representation or expression, as they must be, in accordance with the authors' and the present view, or *principle*. This principle entails the necessity of establishing a rigorous mathematical expression for the physical architecture considered or indeed the fundamental physical principles of quantum theory. As they say: "The rules summarized in this section define the operational language of circuits, which has been discussed in detail in a series of inspiring works by Coecke" [9, p. 4]. As B. Coecke comments:

> The underlying mathematical foundation of this high-level diagrammatic formalism relies on so-called *monoidal categories*, a product of a fairly recent development in mathematics. Its logical underpinning is *linear logic*, an even more recent product of research in logic and computer science. These monoidal categories do not only provide a natural foundation for physical theories, but also for proof theory, logic, programming languages, biology, cooking … So the challenge is to discover the necessary additional pieces of structure that allow us to predict genuine quantum phenomena. These additional pieces of structure represent the capabilities nature has provided us with to manipulate entities subject to the laws of quantum theory. [52, p. 1]

---

[22] I think, that, in accordance with the definition given at the outset, "postulate" may be a better term, because one can hardly have self-evidence of "axioms," but this is a secondary matter, which, as I said, does not really affect the essence of the situation.

This may indeed be a more natural way to give the fundamental structures and principles of quantum theory a proper mathematical expression. As noted above, the authors need five additional axioms for their derivation of the finite-dimensional quantum theory. As they say:

> In addition to the purification postulate, our derivation of quantum theory is based on five informational axioms. The reason why we call them "axioms," as opposed to the purification "postulate," is that they are not at all specific of [to?] quantum theory. These axioms represent standard features of information processing that everyone would, more or less implicitly, assume. They define a class of theories of information processing that includes, for example, classical information theory, quantum information theory, and quantum theory with superselection rules. The question whether there are other theories satisfying our five axioms and, in case of a positive answer, the full classification of these theories is currently an open problem. Here we informally illustrate the five axioms, leaving the detailed description to the remaining part of the paper:
>
> (1) *Causality*: the probability of a measurement outcome at a certain time does not depend on the choice of measurements that will be performed later.[23]
>
> (2) *Perfect distinguishability*: if a state is not completely mixed (i.e., if it cannot be obtained as a mixture from any other state), then there exists at least one state that can be perfectly distinguished from it.
>
> (3) *Ideal compression*: every source of information can be encoded in a suitable physical system in a lossless and maximally efficient fashion. Here *lossless* means that the information can be decoded without errors and *maximally efficient* means that every state of the encoding system represents a state in the information source.
>
> (4) *Local distinguishability*: if two states of a composite system are different, then we can distinguish between them from the statistics of local measurements on the component systems.
>
> (5) *Pure conditioning*: if a pure state of system AB undergoes an atomic measurement on system A, then each outcome of the measurement induces a pure state on system B. (Here *atomic measurement* means a measurement that cannot be obtained as a coarse graining of another measurement.) [9, p. 3]

Importantly, "all these axioms are satisfied by classical information theory" [9, p. 3]. The authors also "make precise the usage of the expression 'operational principle' in the context of [their] paper," which point should not be missed if one wants to properly understand their argumentation, as grounded in fundamental physical (informational) principles:

> By [an] operational principle we mean here a principle that can be stated using only the operational-probablistic language, i.e., using only
>
> (1) the notions of system, test, outcome, probability, state, effect, transformation;
>
> (2) their specifications: atomic, pure, mixed, completely mixed; and
>
> (3) more complex notions constructed from the above terms (e.g., the notion of "reversible transformation").
>
> The distinction between operational principles and principles referring to abstract mathematical properties, mentioned in the Introduction, should now be clear: for example, a statement like "the pure states of a system cannot be cloned" is a valid operational principle, because it can be analyzed in basic operational-probabilistic terms as 'for every system A there exists no transformation C with input system A and output system AA such that C |φ⟩ = |φ⟩|φ⟩ for every pure state φ of A.' On the contrary [By contrast?], a statement like 'the state space of a system with two perfectly distinguishable states is a three-dimensional sphere' is not a valid operational principle, because there is no way to express what it means for a state space to be a three-dimensional sphere in terms of basic operational notions. The fact that a state [space] is a sphere may be eventually derived from operational principles, but cannot be assumed as a starting point. [9, p. 6]

---

[23] As explained earlier (Note 6), this principle is different from that of classical causality (indeed already by virtue of the principle's appeal to probability), while being consistent with relativity.

This distinction is indeed essential, even though operational principles must be given a proper mathematical expression in the formalism of the theory.

The (principle) way of thinking just outlined is equally central to D'Ariano and Perinotti's derivation of Dirac's equation, in the framework of quantum cellular automata. To cite the authors' abstract:

> Without using the relativity principle, we show how the Dirac equation in three space dimensions emerges from the large-scale dynamics of the minimal nontrivial quantum cellular automaton satisfying unitarity, locality, homogeneity, and discrete isotropy. The Dirac equation is recovered for small wave-vector and inertial mass, whereas Lorentz covariance is distorted in the ultrarelativistic limit. The automaton can thus be regarded as a theory unifying scales from Planck to Fermi. [11, p. 1]

As I said, I cannot address the concept of cellular quantum automaton, or the technical mathematical aspects of this derivation, which cannot be given here the treatment they deserve. The two main points that I want to emphasize are as follows. The first is the significance of the principles of information processing and of quantum theory, or even of fundamental physics in general, and establishing what they are, and indeed developing the understanding of these principles. This task is far from being completed, and in some respects it has barely begun to be undertaken. The fact, however, that the authors are able to give their principles a proper mathematical expression is important, and doing so is a significant accomplishment. The second point is, again, that Dirac's equation could be derived by using principles of information processing alone without using the relativity principle. Although this derivation is only a starting point as far as high-energy physics, governed by quantum field theory, is concerned, it is clear that it has major implications for foundational thinking in fundamental physics, including when it comes to large-scale physics.

As noted above, "the Dirac equation in three space dimensions *emerges from the large-scale dynamics* of the minimum-dimension quantum cellular automaton [QCA], satisfying [linearity], unitarity, locality, homogeneity, and discrete isotropy of interactions [without appealing to special relativity]. The Dirac equation is recovered for small wave vector and inertial mass, whereas Lorentz covariance is generally distorted in the ultrarelativistic limit [of very large wave-vectors] [53-56]" [11, p. 1]. This scale is beyond "the usual Fermi scale of high energy physics," because "the QCA as a microscopic mechanism for the emergent quantum field" is proposed "as a framework to unify a *hypothetical* discrete Planck scale with the usual Fermi scale of high-energy physics" [11, p. 1; emphasis added]. The hypothetical nature of some of the assumptions made and the theories alluded to here must be kept in mind, and I shall return to this aspect of the situation presently.

Dirac's equation might be seen as an enactment of the mathematical correspondence principle applied at a level beyond the Fermi scale.[24] This enactment is complex, which should not come as a surprise. Recall that even establishing, via Pauli's spin theory, that establishing Schrödinger's equation as the quantum-mechanical limit of Dirac's equation was nontrivial. Here one deals with a deeper level of, let us say, quantum *reality*, or possibly a reality beyond the quantum, although D'Ariano and Perinotti's framework suggests, at least to this author, that at stake is a certain general quantumness, reflected in linearity and unitarity of their framework at the large scale. The key principles of quantum theory and their quantum-theoretical mathematical expressions and consequences are clearly at work throughout—Hilbert spaces, operators, noncommutativity, the tensor structure, symmetry, group representation, and so forth [11, pp. 2-3]. There is some important new mathematics as well, such as that of "quasi-isometric embedding," a profound recent development in geometry, primarily due to M. Gromov and his geometric-group theory [11, p. 3]. The QCA is, then, the *quantum* cellular automaton operative at the large scale. The derivation of Dirac's equation is achieved by finding Dirac automata, which are "obtained by locally coupling Weyl automata" and give the Dirac equation at the relativistic limit from the informational principles proposed, *all of these principles*, rather than only linearity and unitarity, which are quantum in character [11, pp. 5, 6-9]. This is a reflection of the fact that Dirac's spinors are composed of Weyl's spinors.

---

[24] On the other hand, the article provides "an analytical description of the QCA for the narrow-band states of quantum field theory in terms of a dispersive Schrödinger equation holding at all scales" [11, pp. 1, 4].

In this regard, the situation is parallel to deriving the finite-dimensional quantum theory from a larger (than only quantum) set of principles. Locality, homogeneity, and isotropy are additional principles here, which allow one to dispense with the principle of relativistic (Lorenz) invariance in deriving the equation. Locality merits a special attention here, both in view of its arguably primary role in this derivation and in general. The locality requirement itself and the corresponding principle are defined by the requirement that the cardinality of the set, $S_g$, of sites $g'$, interacting with the site $g$ (both from a denumerable set $G$ of systems, involved in the cellular automaton and described by Fermionic field operators) is "uniformly bound over $G$, namely, $|S_g| \leq k < \infty$ for every $g$" [11, p. 2]. This locality requirement or principle is a more general and deeper conception than the concept of locality associated with the Lorentz invariance, which reemerges from this principle at the corresponding (relativistic) limit and the standard (Fermi) scale of quantum field theory, and hence of Dirac's equation.[25] The significance of the locality requirement is, however, more general and is critical to the informational QCA framework proposed by the authors. In particular, according to D'Ariano (private communication), while falsifiability is a crucial (Popperian-like) requirement to be satisfied by this framework, it is equally crucial that all falsifiability is *local*, which may indeed be seen as a fundamental principle in its own right.

The QCA as such is, again, an enactment of large-scale quantum dynamics, the quantum character of which is manifest in and is defined by linearity and unitarity. There are significant potential implications and benefits of this broader conception beyond the fact the Dirac equation is the Fermi-scale limit of the theory:

> The additional bonus of the automaton is that it also represents the canonical solution to practically all issues raised by both the continuum and the infinite-volume of the field description, such as all divergencies and the problem of particle localizability, all due to the continuum, infinite volume, and the Hamiltonian description. Moreover the QCA is the ideal framework for a quantum theory of gravity, being quantum *ab initio* (the QCA is not derivable by quantizing a classical theory), and naturally incorporates the informational foundation for the holographic principle, a relevant feature of string theories [57, 58] and the main ingredient of the microscopic theories of gravity of Jacobson [59] and Verlinde [60]. Finally, a theory based on a QCA assumes no background, but only interacting quantum systems, with space-time and mechanics as emergent phenomena.
>
> The assumption of Planck-Scale discreteness has the consequence of breaking Lorentz covariance along with all continuous symmetries: These are recovered at the Fermi scale in the same way as in the doubly-special relativity of Amelino-Camelia [53, 54], and in the deformed Lorentz symmetry of Smolin and Magueijo [55, 56]. Such Lorentz deformations have phenomenological consequences, and possible experimental tests have been recently proposed by several authors [61-64]. The deformed Lorentz group of the automaton has been preliminarily analyzed in Ref. [65]. [11, p. 1]

---

[25] This concept has further implications for our understanding of the EPR-type experiments and related problematics, as well as for the question of causality, because relativity (both special and general) is a classically causal theory, while quantum theory, including as thus extended beyond the Fermi scale, is not. It is only causal relativistically or, again, locally in the sense here defined. These subjects would, however, require separate treatments, which cannot be undertaken here.

It is far beyond my scope to address these aspects of the program and their implications, which are exciting and promising but which would also require a careful analysis, including considering the works cited here. However, if the potential significance of the proposed program for large-scale physics is thus made apparent, the hypothetical nature of the theories mentioned here should be kept in mind as well. To cite D'Ariano and Perinotti' conclusion:

> In conclusion, we remark that [the] Lorentz covariance is obeyed only in the relativistic limit $|\mathbf{k}| \ll 1$, whereas the general covariance (corresponding to invariance of $\omega_{\mathbf{k}}^{E\pm}$) is a nonlinear deformation of the Lorentz group, with additional invariants in the form of energy and distance scales [65], as in the doubly-special relativity [53, 54] and in the deformed Lorentz symmetry [55, 56], for which the automaton then represents a concrete microscopic theory. Correspondingly, also [the] CPT symmetry of Dirac's QCA is broken in the ultrarelativistic scale. [11, p. 9]

The theories invoked here remain highly hypothetical. Amelino-Camelia and Piran, cited here, call a hypothetical suspension of the (Lorentz) relativistic invariance a "drastic assumption," albeit made to address certain puzzling data [54, p. 1]. So, it may pay off to be cautious, even though and indeed because alternative proposals, some of which are based on alternative theories of cellular automata [e.g., 66], are equally hypothetical.[26] However, that Dirac's equation could be derived without using the relativity (Lorentz invariance) principle, in addition to being a major achievement in its own right, manifests a promise and potential of the authors' informational program for exploring fundamental physics on scales beyond those of quantum field theory.

Profound foundational questions are at stake. How these questions will be pursued and what role the principles of quantum informational will play in it is difficult to predict. The present article does not aim to do so. It does suggest, however, that fundamental physical principles may play a more significant role than they have more recently in approaching these questions. Equally crucially and perhaps more radically, it also suggests, with the help of both Dirac's own and D'Ariano and Perinotti's derivations of Dirac's equation, that the fundamental principles of *quantum physics* might prove to be more crucial in this pursuit than those of relativity, not inconceivably, general relativity included. In other words, *quantum principles* may also prove to be more crucial rather than those of general relativity in understanding gravity, although, as explained above, some form of locality principle may be equally crucial, and homogeneity and isotropy are likely to be required as well, as very general principles. In other words, while relativity may be abandoned on the large scale (extending to the Planck scale), quantum theory is likely to survive, reflecting the ultimately quantum character of nature. If so, and one cannot, once again, be certain, we may need yet new quantum principles, possibly conceived on lines of principles of information processing and quantum cellular automata, even if developed, as these informational principles are, by building on some among the older principles. But then, this may be our best way to arrive at new principles in physics or elsewhere.

**Acknowledgements.** I am grateful to G. Mauro D'Ariano for sharing, in many invaluable discussions, his thinking and his knowledge of quantum theory. I would also like to thank Lucien Hardy, Gregg Jaeger, Andrei Khrennikov, and Paolo Perinotti for productive exchanges that helped my work on this article. I would like to add that the authors mentioned here, as well as the present author, have each published a series of papers on quantum foundations in the Proceedings of Växjö conferences on quantum foundations during the last decade. I gratefully acknowledge the role of these conferences in my work.

---

[26] For yet another type of alternative approach, see [67].